\documentstyle[12pt,psfig]{article} \def\inbar{\,\vrule height 1.5ex width
.4pt depth0pt} \def\IB{\relax{\rm I\kern-.18em B}}
\def\IC{\relax\hbox{$\inbar\kern-.3em{\rm C}$}} \def\ID{\relax{\rm
I\kern-.18em D}} \def\IE{\relax{\rm I\kern-.18em E}} \def\IF{\relax{\rm
I\kern-.18em F}} \def\IG{\relax\hbox{$\inbar\kern-.3em{\rm G}$}}
\def\IH{\relax{\rm I\kern-.18em H}} \def\II{\relax{\rm I\kern-.18em I}}
\def\IK{\relax{\rm I\kern-.18em K}} \def\IL{\relax{\rm I\kern-.18em L}}
\def\IM{\relax{\rm I\kern-.18em M}} \def\IN{\relax{\rm I\kern-.18em N}}
\def\IO{\relax\hbox{$\inbar\kern-.3em{\rm O}$}} \def\IP{\relax{\rm
I\kern-.18em P}} \def\IQ{\relax\hbox{$\inbar\kern-.3em{\rm Q}$}}
\def\IR{\relax{\rm I\kern-.18em R}} \def\IZ{\relax\ifmmode\lrefhchoice
{\hbox{\cmss Z\kern-.4em Z}}{\hbox{\cmss Z\kern-.4em Z}}
{\lower.9pt\hbox{\cmsss Z\kern-.4em Z}} {\lower1.2pt\hbox{\cmsss
Z\kern-.4em Z}}\else{\cmss Z\kern-.4em Z}\fi} \font\cmss=cmss10
\font\cmsss=cmss10 at 7pt \def\IZ{\relax\ifmmode\mathchoice {\hbox{\cmss
Z\kern-.4em Z}}{\hbox{\cmss Z\kern-.4em Z}} {\lower.9pt\hbox{\cmsss
Z\kern-.4em Z}} {\lower1.2pt\hbox{\cmsss Z\kern-.4em Z}}\else{\cmss
Z\kern-.4em Z}\fi} \def\H{{\cal H}} \def\dr{\D^R} 
 \def\g{\gamma_4} \def\e{\varepsilon} \def\vp{\varphi}
\def\th{\theta}  \def\half{{1\over2}} \def\semi{\sim}
\def\m{{\cal M}_{0,4}} \def\ds{\displaystyle} \def\d{{\rm d}} \def\D{{\cal
D}_{0,4}} \def\dgn{{\cal D}_{G,N}} \def\pgn{\widehat{\cal P}_{G,N}}
\def\JS{\D^{R}}  \def\o{\omega} \def\n{\eta}
\def\z{\zeta}  \def\l{\lambda} \def\lz{\l^{\rm zeros}}
\def\lp{\l^{\rm poles}} \def\SB{\JS} \def\im{\Im{\rm m}} \def\re{\Re{\rm
e}} \def\a{\alpha} \def\to{\rightarrow} \def\b{\beta} \def\qt{q_\tau}
\def\O{{\rm O}} \def\V{{\cal V}_{0,4}}
\def\exch{\leftrightarrow}  
 \def\\#1{_{{}_{#1}}} \def\v{{\rm v}} \def\mgn{{\cal
M}_{G, N}} \def\V{{\cal V}} \def\mon{{\cal M}_{0, N}} \def\m{{\cal M}_{0,
4}} \def\lf{\left} \def\rt{\right} \def\s{\sigma} \def\qt{q_\tau}
\def\qu{q_u} \def\quO{q_{u_0}}  \def\dl{\delta} \def\res{{\rm
Res}} \def\qa{q_\a} \def\to{\rightarrow} \def\S{\Sigma} 
\def\F{{\cal F}} \def\f{\F_{0,4}} \def\gO{\gamma^{(0)}_4}

\def\massive{\hbox{\lower4pt\vbox{\hbox{\kern+10pt\hbox{$\scriptstyle X$}}
      \kern-3pt \hbox{\psfig{figure=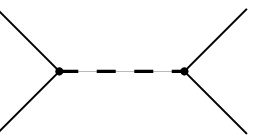,height=12pt}}}}}
\def\vfour{\hbox{\lower4pt\psfig{figure=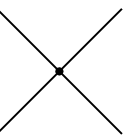,height=12pt}}}
\def\tach{\hbox{\lower4pt\vbox{\hbox{\kern+10pt\hbox{$\scriptstyle\tau$}}
      \kern-3pt \hbox{\psfig{figure=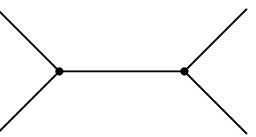,height=12pt}}}}}
\def\full{\hbox{\lower4pt\psfig{figure=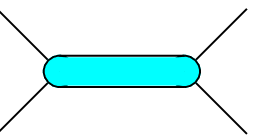,height=12pt}}}
\def\tachv{{\scriptstyle\v_3}\kern-3pt\tach\kern-3pt{\scriptstyle\v_3}}

\title{{}\hfill \vbox{\hbox{\small MIT-CTP 2388} \hbox{\small\tt
                         hep-th/9412106}}\\[0.5in] Effective Tachyonic
                         Potential in Closed String Field Theory}
                         \author{Alexander Belopolsky\thanks{ E-mail
                         address: {\tt belopols@ctpa02.mit.edu}\hfill\break
                         This work is supported in part by funds provided
                         by the U.S.  Department of Energy (D.O.E.)  under
                         cooperative agreement \#DF-FC02-94ER40818.}\\
                         Center for Theoretical Physics\\ MIT}
                       \date{December 12, 1994}
\begin{document}
%---TeX-end-of-header---

\maketitle
\begin{abstract}
  We calculate the effective tachyonic potential in closed string field
  theory up to the quartic term in the tree approximation. This involves an
  elementary four-tachyon vertex and a sum over the infinite number of
  Feynman graphs with an intermediate massive state. We show that both the
  elementary term and the sum can be evaluated as integrals of some measure
  over different regions in the moduli space of four-punctured spheres.  We
  show that both elementary and effective coupling give negative
  contributions to the quartic term in the tachyon potential.  Numerical
  calculations show that the fourth order term is big enough to destroy a
  local minimum which exists in the third order approximation.
\end{abstract}

\section{Introduction}
\label{Intr}
Bosonic closed string field theory (CSFT) has been formulated as a full
quantum field theory in Ref.~\cite{zwiebachlong}. It was shown to be
locally background independent in Refs.~\cite{zwie-qua,asen-pro,asen-bac}.
Currently there is no manifestly background independent formulation of the
CSFT available.  To formulate a CSFT we have to specify a conformal field
theory (CFT) and then construct a closed string action as a functional on
the state space $\H$ of the CFT. This action should satisfy the BV master
equation \cite{zwiebachlong}. Given a CFT there are many different choices
for the master action. One possibility stemming from minimal area metrics
was described in Refs.~\cite{zwie-mi1,zwie-mi2,zwie-mi3}. This action has
an interesting property: it minimizes the tachyonic potential order by
order in perturbation theory \cite{bz}.

The following expression for the classical tachyonic potential has been
obtained by G.~Moore \cite{moore} and was proven in Ref.~\cite{bz}:
\begin{equation}
V (t)= -t^2-\sum_{N=3}^\infty \,\v\\N\, {t^N\over N!}\, ,\quad
\label{pot}
\end{equation}
where
\begin{equation}
\v\\N \equiv (-)^N{2\over\pi^{N-3}}\, \int_{\V_{0,N}}
\bigl(\prod_{I=1}^{N-3} {\d}^2 \xi\\I\,\bigr) \prod_{I=1}^N \lf|{
w'\\I(\xi\\I)}\rt|^2.
  \label{v-tach}
\end{equation}
The global uniformizer $\xi$ is chosen such that the coordinates of the
last three punctures are $\xi\\{N-2} = 0$, $\xi\\{N-1} = 1$ and $\xi\\{N} =
\infty$. $w\\I(\xi)$ denotes the local coordinate around the $I$-th
puncture and the derivative at infinity is to be taken with respect to
$1/\xi$. The integration in (\ref{v-tach}) has to be performed over
$\V_{0,N}$, the region of the moduli space which can not be covered by the
string diagrams with a propagator. We will distinguish the {\em missing
region} or the {\em string vertex} $\V_{0,N}$ from the {\em Feynman region}
$\F_{0,N}=\mon\backslash\V_{0,N}$.

The cubic term does not require integration and can be easily evaluated
(see Refs.~\cite{kost-con,bz}).
\begin{equation}
\v_3 = -{3^9\over 2^{11}} \approx -9.61.
  \label{v3}
\end{equation}
For $N>3$ there are two major obstacles to evaluation of (\ref{v-tach}):
firstly, we need a description of $\V\\{0,N}$ and secondly we have to
define the local coordinates $w\\I$.  Unfortunately, the string field
theory defines the vertex and the local coordinates implicitly in terms of
a quadratic differential of special type and its invariants. Finding the
quadratic differential is a difficult problem on its own and even when an
analytic expression for it is known to find the desired invariants is still
not trivial.  In this article we will deal mostly with the fourth order
term:
\begin{equation}
 \v_4 \equiv {2\over\pi}\,\int_{\V_{0,4}}\mu,
  \label{v4}
\end{equation}
where $\mu$ is the measure of integration on the moduli space. It can be
expressed in terms of local coordinates $w_i(\xi)$ as
\begin{equation}
  \label{mu}
\mu= \left|w_1'(\l)\,w_2'(0)\,w_3'(1)\,w_4'(\infty)\right|^2\d^2\l.
\end{equation}
As before the global uniformizer $\xi$ is fixed by placing three punctures
at $\xi=0$, $1$ and $\infty$. The coordinate of the fourth puncture $\l$
provides a coordinate on the moduli space $\m$.  We will use a notation
$\d^2\l$ to denote the standard measure $\d\re\,\l\,\d\im\,\l$ on the
complex plane of $\l$.

\noindent \underline{Effective potential.} The bare tachyonic potential
defined by (\ref{pot}) and (\ref{v-tach}) is not a physical quantity
because the tachyon is coupled to the other fields in the string field
theory.  In order to calculate an effective potential (which is physical)
one has to perform a summation over all the diagrams with intermediate
non-tachyon states.  Thus the effective four tachyon coupling constant
$\v_4^{\rm eff}$ consists, in the tree approximation, of the elementary
coupling $\v_4=\vfour$ and the sum over infinite number of diagrams with
intermediate massive states $X$. We can write it schematically as
\begin{equation}
\v_4^{\rm eff}=\vfour+\sum_X\massive.
\label{v4effd}
\end{equation}
Instead of summing over all massive states we will calculate the full sum
over all the states including the tachyon as an integral over the Feynman
region~$\f=\m\backslash\V_4$
$$ \full=\int_{\f}\mu=\tach+\sum_X\massive.
$$ The first term with an intermediate tachyon can be easily evaluated in
terms of the three-tachyon coupling constant $\v_3$:
\begin{equation}
\tachv = 3\cdot{\v_3^2\over p^2+m^2}=-{3\over2}\,\v_3^2,
\end{equation}
where $p=0$ is the momentum of a propagating tachyon and $m_\tau^2=-2$ is
its mass squared.  The factor of three comes from the sum over three
channels each giving the same contribution.  Combining the above equations
we find
\begin{equation}
\v_4^{\rm eff}=\vfour+\full-\tach= \v_4+{2\over\pi}\int_{\f}\mu +
{3\over2}\,\v_3^2.
\label{eff}
\end{equation}
We will see that the integral in (\ref{eff}) is divergent and has to be
found by analytic continuation.

For the case of $\m$ the invariants of the quadratic differential can be
expressed in terms of elliptic integrals.  Our discussion will involve an
extensive use of elliptic functions and their $q$-expansions. These
$q$-expansions prove to be a powerful tool in numerical calculations.

The paper is organized as follows. First of all we will derive a general
formula for the four-tachyon amplitude. We express the amplitudes in terms
of invariants ($\chi_{ij}$) of a four-punctured sphere with a choice of
local coordinates.  Then in sect.~\ref{Howa} we will review some basic
properties of quadratic differentials and show how a quadratic differential
defines local coordinates in general. In sect.~\ref{Quad} we will apply the
general construction of sect.~\ref{Howa} to the case of $\m$. We introduce
integral invariants $a$, $b$ and $c$ associated with a quadratic
differential with four second order poles. In sect.~\ref{Miss} we show that
the integrals over $\V_{0,4}$ and $\F_{0,4}$ can be easily evaluated if we
know the integrand in terms of $a$ and $b$. In
sects.~\ref{Main},~\ref{Infi} and~\ref{From} we express the measure of
integration as a function of $a$ and $b$. We reduce the problem to a single
equation involving elliptic functions, which we solve approximately in two
limits: one corresponding to a long propagator and an arbitrary twist angle
and the other corresponding to both propagator and twist being small. For
the intermediate region we solve the equation numerically.  Finally we
calculate the contribution of the Feynman diagrams (\ref{eff}) in
sect.~\ref{Summ} and the elementary coupling (\ref{v4}) in
sect.~\ref{Bare}.

\section{Four tachyon off-shell amplitude}
\label{Four}
\setcounter{equation}{0} In this section we will derive a formula for the
scattering amplitude of four tachyons with arbitrary momenta. Although for
the tachyonic potential we only need the amplitude at zero momentum, the
integral which defines it is divergent and we are forced to treat it as an
analytic continuation from the region in the momentum space where it
converges. We will give the details on the origin of this divergence in
sect.~\ref{Summ}.

A general formula for the tree level off-shell amplitude has been found in
Ref.~\cite{bz} and for the case of four tachyons it gives the off-shell
Koba-Nielsen formula
\begin{equation}
\Gamma_4(p_1,p_2,p_3,p_4) = {2\over\pi}\int{\d^2\l\over|\l(1-\l)|^2}\;
      |\chi_{1234}|^2\cdot\prod_{i<j}|\chi_{ij}|^{p_ip_j},
  \label{4tachp}
\end{equation}
which expresses the four tachyon amplitude in terms of $\psl$ invariants
$\l$, $\chi_{ij}$ and $\chi_{1234}$. The first invariant is just the cross
ratio of the poles which we define~as\footnote{Here we use a different
cross ratio to that in Ref.~\cite{bz}. In order to use the formulae of
Ref.~\cite{bz} one has to change $\l$ to $\l/(1-\l)$}
\begin{equation}
\l={z_1-z_2\over z_1-z_3}\cdot{z_3-z_4\over z_2-z_4}.
  \label{crossr}
\end{equation}
The $\chi$ invariants can be expressed in terms of the mapping radii
$\rho_i$ as
\begin{equation}
\chi_{ij}={(z_i-z_j)^2\over\rho_i\rho_j}.
  \label{chiij}
\end{equation}
Unlike those in Ref.~\cite{bz} the $\chi$ invariants and mapping radii used
here are complex numbers.  We achieve the complexification by keeping the
phases of the local coordinates. Thus, here $\rho_i$ is given by
$$ \rho_i={1\over w_i'(z_i)},
$$ and not just the absolute value. The last invariant $\chi_{1234}$ can be
expressed in terms of $\chi_{ij}$ as
\begin{equation}
\chi_{1234}^2 = \chi_{12}\chi_{23}\chi_{34}\chi_{41},
  \label{1234}
\end{equation}
By definition $\chi_{ij}=\chi_{ji}$ and thus for a four punctured sphere we
have ${4\choose2}=6$ different invariants. We will call a choice of local
coordinates symmetric if the local coordinates do not change under the
symmetries of the Riemann surface. Specifically, if $S$ is an automorphism
of a punctured Riemann surface $\Sigma$ which maps the $i$-th puncture to
the $j$-th puncture, we require that
\begin{equation}
  \label{recdef}
  w_j(S(\sigma))=w_i(\sigma),
\end{equation}
where $\sigma\in\Sigma$ belongs to the $i$-th coordinate patch. It is well
known, that in most cases this condition can only be satisfied up to a
phase (see Ref.~\cite{zwie-cov}). Nevertheless, for a general
four-punctured sphere the phases can be retained. Four-punctured spheres
have a unique property: there exists a non-trivial symmetry group which
acts on {\em any\/} four-punctured sphere. This group consists of the
automorphisms which interchange two distinct pairs of punctures. One can
easily check that these automorphisms exist for any $\Sigma\in\m$.  One can
visualize this symmetry by placing the punctures at the vertices of a
rectangle --- the symmetry group then becomes the group of the rectangle
$\IZ_2\times\IZ_2$. There are a couple of four-punctured spheres which have
a larger symmetry group: a tetrahedral symmetry in the case of $\l=\exp(\pi
i/3)$, which is the most symmetric case or the symmetry group of the square
for $\l=-1,\,1/2$, or $2$. It is not possible to realize the symmetry
conditions for these larger groups if we wish to retain the phases,
therefore we can require that (\ref{recdef}) holds only for
$S\in\IZ_2\times\IZ_2$.

For symmetric local coordinates the six $\chi$-invariants are not
independent. Using $\IZ_2\times\IZ_2$ symmetry one can prove that
\begin{equation}
  \begin{array}{rclrcl}
    \chi_{12}&=&\chi_{34}&\equiv&\chi_s,\\[12pt]
    \chi_{14}&=&\chi_{23}&\equiv&\chi_t,\\[12pt]
    \chi_{13}&=&\chi_{24}&\equiv&\chi_u.\\[12pt]
  \end{array}
  \label{recsym}
\end{equation}
Furthermore, due to the transformation properties of the mapping radii
\begin{equation}
  \label{chi-lam}
  \chi_s/\chi_u=-\l,\qquad\mbox{and}\qquad\chi_t/\chi_u=\l-1,
\end{equation}
and thus
\begin{equation}
\chi_u+\chi_s+\chi_t=0.
  \label{s+t+u}
\end{equation}
Equations (\ref{recsym}) and (\ref{s+t+u}) show that for a symmetric choice
of local coordinates there are only two independent $\chi$-invariants.

Now we can rewrite the Koba-Nielsen formula in terms of $\chi_s$, $\chi_t$,
$\chi_u$ and the Mandelstam variables
\begin{equation}
\Gamma_4(s,t,u) =
{2\over\pi}\int{|\chi_s\chi_t|^2\d^2\l\over|\l(1-\l)|^2}\; |\chi_s|^{t+u-s}
|\chi_t|^{u+s-t} |\chi_u|^{s+t-u}.
  \label{kobastu}
\end{equation}
Note that the momentum dependent part of (\ref{kobastu}) is manifestly
symmetric with respect to $s$, $t$ and $u$. Let us show that the momentum
independent part is symmetric as well. First of all we introduce a
differential one-form
\begin{equation}
\g(s,t,u)={\chi_s\chi_t\d\l\over\l(1-\l)} \chi_s^{t+u-s\over2}
\chi_t^{u+s-t\over2} \chi_u^{s+t-u\over2}.
  \label{gamma}
\end{equation}
Given a differential one-form $\o=\o(\l)\d\l$ we can define the
corresponding measure as $\mu = |\o|^2=|\o(\l)|^2\d^2\l$.  The measure of
integration in (\ref{kobastu}) is just $|\g(s,\,t,\,u)|^2$ and we rewrite
the Koba-Nielsen formula as
\begin{equation}
\Gamma_4(s,\,t,\,u)={2\over\pi}\int|\g(s,\,t,\,u)|^2.
  \label{Gg}
\end{equation}
Consider the momentum independent part of $\g$:
\begin{equation}
\gO=\g(0,0,0) ={\chi_s\chi_t\d\l\over\l(1-\l)}= \chi_s\d\chi_t -
 \chi_t\d\chi_s,
  \label{gst}
\end{equation}
where we have made use of~(\ref{chi-lam}). We can now use
$\chi_u+\chi_s+\chi_t=0$ and show that
\begin{equation}
\gO = \chi_t\d\chi_u-\chi_u\d\chi_t = \chi_u\d\chi_s - \chi_s\d\chi_u,
  \label{gut}
\end{equation}
and hence that $|\gO|^2$ is totally symmetric.

The following expression for $\gO$ although not explicitly symmetric is
very simple and will be particularly useful latter. Using~\ref{chi-lam} we
can rewrite~\ref{gst} as
\begin{equation}
\gO=\chi_u^2\d\l.
  \label{useg}
\end{equation}

In the spirit of the string field theory we distinguish the contribution
from the Feynman region $\F_{0,4}\subset\m$ (the surfaces which can be sewn
out of two Witten's vertices and a propagator) and the missing region
$\V_{0,4}=\m\backslash\F_{0,4}$. The later appears in the string field
theory as the elementary four tachyon coupling
\begin{equation}
  \v_4={2\over\pi}\int_{\V_{0,4}}|\chi_u|^4\d^2\l.
  \label{v4amp}
\end{equation}

\section{How a quadratic differential defines local coordinates.}
\label{Howa}
\setcounter{equation}{0} As we mentioned in the introduction, the
definition of off-shell string amplitudes requires use of local coordinates
around the punctures of a Riemann surface. In this section we describe how
the local coordinates can be specified by a quadratic differential of
special type.

Given a local coordinate in some region of a Riemann surface, a quadratic
differential can be written as $\phi=\varphi(z)(\d z)^2$.  $\varphi(z)$ is
called the `function element' of the quadratic differential. Although the
value of the function element at a particular point does depend on the
choice of the coordinate, its zeros and poles are
coordinate-independent. The second order poles of quadratic differentials
play a similar role to the simple poles of Abelian differentials. The
residue $\res_p\phi$ (the coefficient of the most singular term in the
Laurent expansion of the function element) of a quadratic differential
$\phi$ at a second order pole $p$ is coordinate independent.

Given a Riemann surface $\S\in\mgn$ of genus~$G$ with $N$~punctures we
define the space $\dgn(\S)$ of quadratic differentials with second order
poles at each puncture and the space $\dgn^R(\S)\subset\dgn(\S)$ restricted
by the condition $\res\,\phi=-1$ at every pole.  The space $\dgn(\S)$ is
finite dimensional with $\dim\,\dgn(\S)=3\,G - 3 + 2\,N$.  Furthermore,
$$\dim\,\dgn^R(\S)=\dim\,\dgn(\S) - N = 3\,G - 3 + N$$ is equal to the
dimension of the moduli space $\mgn$. We consider the spaces of quadratic
differentials with $N$ second order poles $\dgn$ and $\dgn^R$ as fiber
bundles over $\mgn$.

With a quadratic differential $\phi$ we associate a contact field $\phi>0$.
The integral lines of this field are called horizontal trajectories.  We
define a critical horizontal trajectory as one which starts at a zero of
the quadratic differential and the critical graph as the set of horizontal
trajectories which start and end at the zeros.

Let $\pgn$ be the moduli space of the genus~$G$ Riemann surfaces with
$N$~punctures and a choice of local coordinate up to a phase around each
puncture.  One can think of $\pgn$ as of a space of surfaces with
$N$~punctures and a closed curve (coordinate curve) drawn around each
puncture. Due to the Riemann mapping theorem, there is a unique (up to
phase) holomorphic map from the interior of a curve to the unit circle,
which takes the puncture to $0$. This map defines a local
coordinate. Keeping this description in mind one can define an embedding
$\Phi:\;\dgn^R\to\pgn$ using the critical graph of a quadratic differential
to define a set of coordinate curves.

We can describe $\Phi$ more explicitly. Let $\phi\in\dgn^R$ be a quadratic
differential. By definition of $\dgn^R$ it has $N$ second order poles with
residue $-1$. Let $p$ be such a pole.  Then, there exists a local
coordinate $w$ in the vicinity of $p$ such that
\begin{equation}
  \phi=-{(\d w)^2\over w^2}.
  \label{local}
\end{equation}
Indeed, let $z$ be some other coordinate and
\begin{equation}
  \phi=\varphi(z)(\d z)^2.
  \label{other}
\end{equation}
We can find $w(z)$ solving the differential equation $i\,\d
w/w=\varphi^{1/2}(z)\d z$. The solution is given by
\begin{equation}
  w(z)=\exp\lf(-i\int_{z_0}^z\varphi^{1/2}(z')\d z'\rt).
  \label{sol}
\end{equation}
The point $z_0$ may be chosen arbitrarily and, so far, the local coordinate
is defined by (\ref{local}) only up to a multiplicative constant.  Moreover
(\ref{local}) does not change when we substitute $1/w$ for $w$, which is
equivalent to the change of sign of the square root in (\ref{sol}). The
latter arbitrariness can be easily fixed by imposing the condition
$w(p)=0$. The inverse map $z=h(w)$ is a holomorphic function of the local
coordinate, which can be analytically continued to a disk of some radius
$r$. We can always rescale $w$ so that $r=1$. this fixes the scale of
$w$. Now we have to show that the coordinate curves corresponding to this
set of local coordinates form the critical graph of the quadratic
differential.  Indeed, the coordinate curve given by $|w|=1$ is a
horizontal trajectory of the quadratic differential which is equal to $(\d
w)^2/w^2$. Let us show that it has at least one zero on it. By definition
$h(w)$ is holomorphic inside the unit disk and can not be analytically
continued to a holomorphic function on a bigger disk. Yet $h'(w)=\d z/\d
w=1/(w(z)\varphi^{1/2}(z))$ and thus $h(w)$ is holomorphic at $w$ unless
$\varphi(h(w))=0$, or $w$ is the coordinate of a zero of $\phi$. We
conclude then, that there is at least one zero on the curve $w(z)=1$.
Finally we can write a closed expression for the local coordinates
associated with the quadratic differential $\phi$:
\begin{equation}
  w(z)=\exp\lf(-i\int_{z_0}^z\sqrt{\phi}\rt),
  \label{loc-w}
\end{equation}
where the sign of the square root is fixed by $\res_{p}\sqrt\phi=i$ and
$z_0$ is a zero of $\phi$. In general for each pole one has to select a
zero to use in (\ref{loc-w}), but for the most interesting case when
critical graph is a polyhedron choosing a different zero alters only the
phase of $w(z)$.

So far a quadratic differential defines the local coordinates, but it is
not itself defined by the underlying Riemann surface because the dimension
of $\dgn$ is twice as big as the dimension of $\mgn$. In order to fix the
quadratic differential we need an extra $3\,G-3+2N$ complex or $6\,G-6+4N$
real conditions. In the next section we will describe these conditions for
the case $G=0$, $N=4$.

\section{Quadratic differentials with four second order poles}
\label{Quad} \nocite{kaku-ano,kaku-sha,kaku-non}
\setcounter{equation}{0} In this section we focus on the case of a
four-punctured sphere, $G=0$ and $N=4$. We define the integral invariants
$a$, $b$ and $c$ of a quadratic differential which control the behavior of
its critical horizontal trajectories. We find explicit formulae for these
invariants in terms of Weierstrass elliptic functions.

Consider a meromorphic quadratic differential on a sphere which one has
four second order poles. Given a uniformizing coordinate $z$ on the sphere
we can write the quadratic differential as
\begin{equation}
\phi={Q(z)\over \ds\prod_{i=1}^4\,(z-z_i)^2}\;(\d z)^2.
  \label{phi}
\end{equation}
In order for $\phi$ to be holomorphic at $z=\infty$ the polynomial $Q$
should be of degree less than or equal to $4$:
\begin{equation}
Q(z) = a_4\,z^4+a_3\,z^3+a_2\,z^2+a_1\,z+a_0.
  \label{Q}
\end{equation}
So far we have a five-dimensional complex linear space $\D=\IC^5$ of
quadratic differentials.  When we restrict ourselves to quadratic
differentials with the residues\footnote{We call the coefficient of the
${(\d z)^2\over(z-z_0)^2}$ in the Laurent expansion of a quadratic
differential near the point $z_0$ the residue of the quadratic
differential. One can easily see that the residue does not depend on the
choice of a local coordinate.} equal $-1$ at every pole we define a
one-dimensional complex affine subspace $\JS\in\D$.  Now we want to
parameterize $\SB$ in such way that coordinates do not depend on the choice
of global uniformizer $z$. The following combinations of the coordinates of
the poles and the zeros are invariant: the cross ratio of the poles,
\begin{equation}
    \lp ={z_1-z_2\over z_1-z_3}\cdot {z_3-z_4 \over z_2-z_4}\label{lp},
\end{equation}
which parameterize the underlying $\m$, and the cross ratio of the zeros,
\begin{equation}
  \lz ={e_1-e_2\over e_1-e_3}\cdot {e_3-e_4 \over e_2-e_4}\label{lz},
\end{equation}
which fixes the quadratic differential. Such a parameterization is
particularly useful because it separates the fibers of $\JS$ in an obvious
way.

Another parameterization can be obtained as follows. Let $\gamma_{ij}$ be a
set of smooth curves connecting $e_i$ and $e_j$ in such a way that they
form a tetrahedron with the poles $z_i$ on the faces.  The integrals
$$ I_{ij}=\int_{\gamma_{ij}}\sqrt{\phi}
$$ are well defined and do not depend on the deformation of
$\gamma_{ij}$. By contour deformation we can show that the integrals along
the opposite edges of the tetrahedron are equal.  Let
\begin{equation}
  \begin{array}{rcccl}
    a&=&I_{12}&=& I_{34},\cr b&=&I_{23}&=& I_{14},\cr c&=&I_{31}&=& I_{24}.
  \end{array}
  \label{abc:def}
\end{equation}
Again, by contour deformation $a+b+c=2\,\pi$ and thus we have only two
independent complex parameters $a$ and $b$ which can be used as coordinates
on $\JS$. So far $\lp$ and $\lz$ are analytic functions of $a$ and
$b$. Note that we propose here a point of view regarding the $a$, $b$,
$c$-parameters differing from that of Ref.~\cite{poly}. In that paper $a$,
$b$ and $c$ were real by definition and provided a real parameterization of
the moduli space $\m$, while here they are complex and parameterize ${\cal
D}_{0,4}^R$. This will be useful to give a unified description of the
Strebel and Feynman regions as we will show later on in sect.~\ref{Miss}.

In general integrals in (\ref{abc:def}) are complete elliptic integrals of
the third kind.  In order to evaluate them we will need the following
lemma.  \vskip12pt
\par\noindent{\bf Reduction Lemma. } {\it Let $\phi$ be a quadratic
differential on the sphere such that in a uniformizing coordinate $z$ it is
given by
$$ \phi={Q(z)\over \ds\prod_{i=1}^4\,(z-z_i)^2}\;(\d z)^2.
$$ where $Q(z)$ is a polynomial of degree four. The square root of $\phi$
defines an Abelian differential on the Riemann surface $\Sigma$ of
$\sqrt{Q(z)}$. Since $Q(z)$ has degree four, $\Sigma$ is a torus. Let the
periods of the torus be $2\,\o_1$ and $2\,\o_2$. The Abelian differential
$\sqrt\phi$ has periods $\o_1$ and $\o_2$ if all the poles of $\phi$ have
equal residues.  } \vskip12pt
\noindent {\bf Proof.} The proof is based on the $\IZ_2\times\IZ_2$
symmetry of the four-punctured sphere. Let us show that a quadratic
differential $\phi\in{\cal D}_{0,4}$ with equal residues is invariant under
these symmetries. It is convenient to fix the uniformizing coordinate $z$
on the sphere so that the zeros of the quadratic differential have
coordinates $\pm 1$ and $\pm k$. Using this coordinate we can write any
quadratic differential $\phi\in{\cal D}_{0,4}$ with equal residues as
\begin{equation}
  \label{qdj}
    \phi=C {(z^2-k^2)(z^2-1)\over(\z^2z^2-k^2)^2(z^2-\z^2)^2}\,(\d z)^2.
\end{equation}
where $\z$ is a position of one of the poles and $C$ is an arbitrary
constant. The symmetry group is generated by two transformations which can
be written as
\begin{equation}
  \label{recproof}
  S_1:\;z\to-z\quad\mbox{and}\quad S_2:\;z\to k/z.
\end{equation}
We can extend this symmetry to the Riemann surface of $\phi$ which is a
torus given by
\begin{equation}
  \label{torwz}
  w^2=(z^2-k^2)(z^2-1).
\end{equation}
The generators $S_k$ act on $w$ by
\begin{equation}
  \label{recw}
  S_1:\;w\to -w\quad\mbox{and}\quad S_2:\;w\to -k\,{w\over z^2}.
\end{equation}
Clearly, (\ref{recw}) together with (\ref{recproof}) define the symmetries
of the torus given by (\ref{torwz}). A holomorphic Abelian differential on
the torus $\d u=\d z/w$ is invariant under these transformations and
therefore $S_k$ are translations of the torus. By definition $S_k^2=1$ and
we conclude that $S_k$ is a translation by half a period,
$S_k(u)=u+\o_k$. The square root of the quadratic differential can be
written in terms of $\d u$ as
\begin{equation}
  \label{sqrtphi}
   \sqrt{\phi}=\sqrt{C}{w^2\d u\over(\z^2z^2-k^2)(z^2-\z^2)}.
\end{equation}
Expression (\ref{sqrtphi}) is invariant under $S_k$ and therefore
$\sqrt{\phi}$ has periods $\o_1$ and $\o_2$. QED.
\begin{figure}[ht]
  \begin{center}
    \leavevmode \psfig{figure=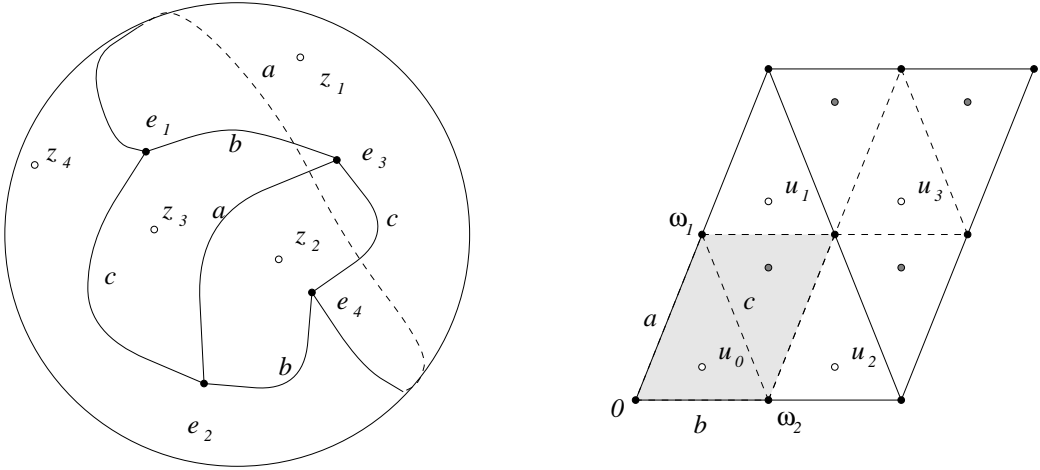,width=\textwidth}
  \end{center}
  \caption{The sphere and the torus.}
  \label{fig:sphere}
\end{figure}
Let $u$ be a coordinate on the torus and $[2\o_1, 2\o_2]$ be its periods.
For a quadratic differential $\phi\in\dr$ the reduction lemma states that
if $\sqrt\phi = f(u)\d u$ then $f(u)$ has periods $[\o_1, \o_2]$. The
quadratic differential has four second order poles with residue $-1$, and
four simple zeros. Thus, $\sqrt\phi$ has eight poles with residue $\pm i$
and four double zeros, or equivalently, $f(u)$ has two poles and a double
zero in its fundamental parallelogram. In Fig.~\ref{fig:sphere} we show the
sphere and the torus with the positions of the poles and zeros marked. The
shaded region on the torus is the fundamental parallelogram of $f(u)$.

\begin{figure}[t]
  \begin{center}
    \leavevmode \setlength{\unitlength}{1.in}
  \begin{picture}(5.2,4)
    \put(0,2){\psfig{figure=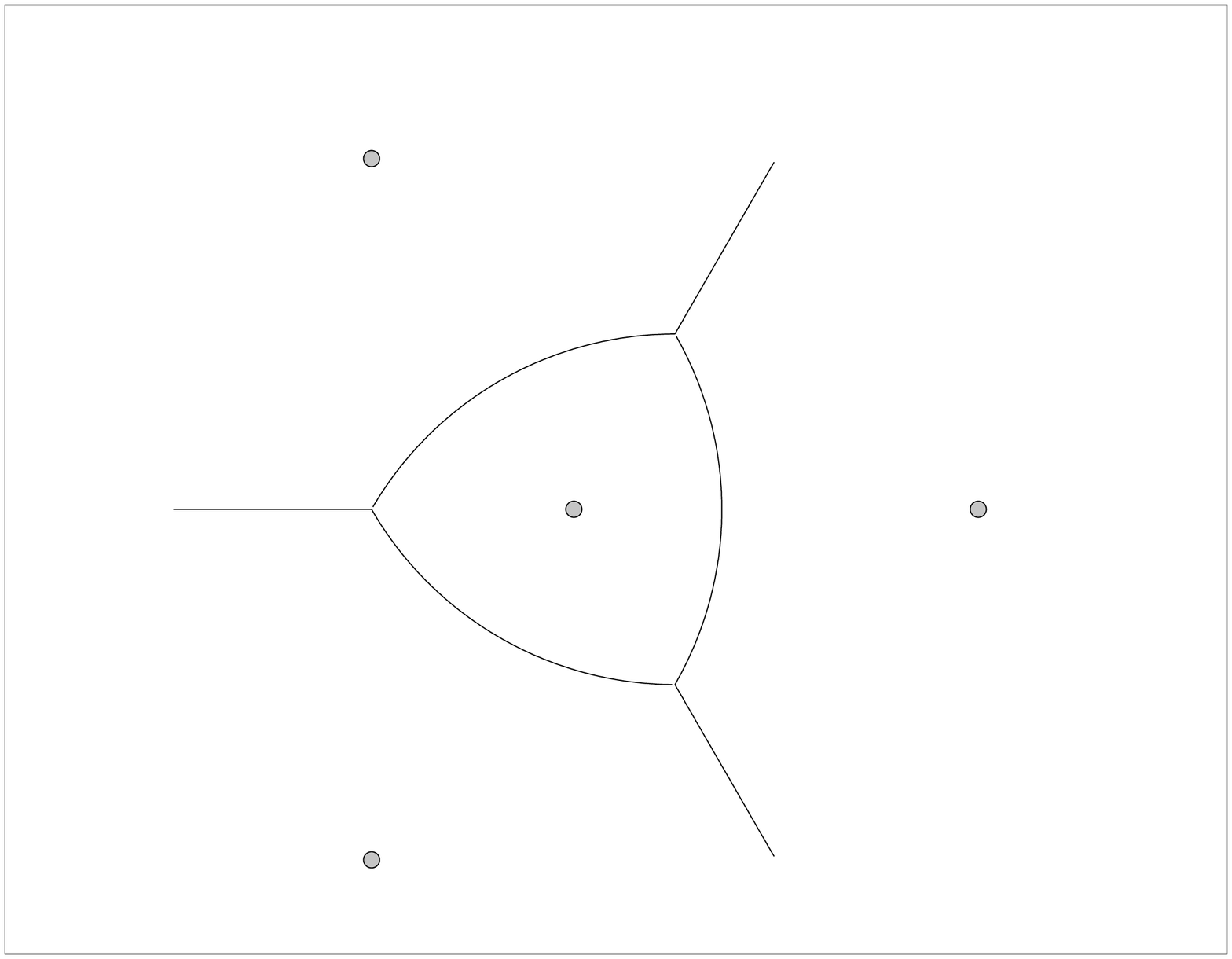,width=2.6in}}
    \put(1.5,3.75){$a=b=2\pi/3$} \put(0.1,3.75){case (1)}
    \put(2.6,2){\psfig{figure=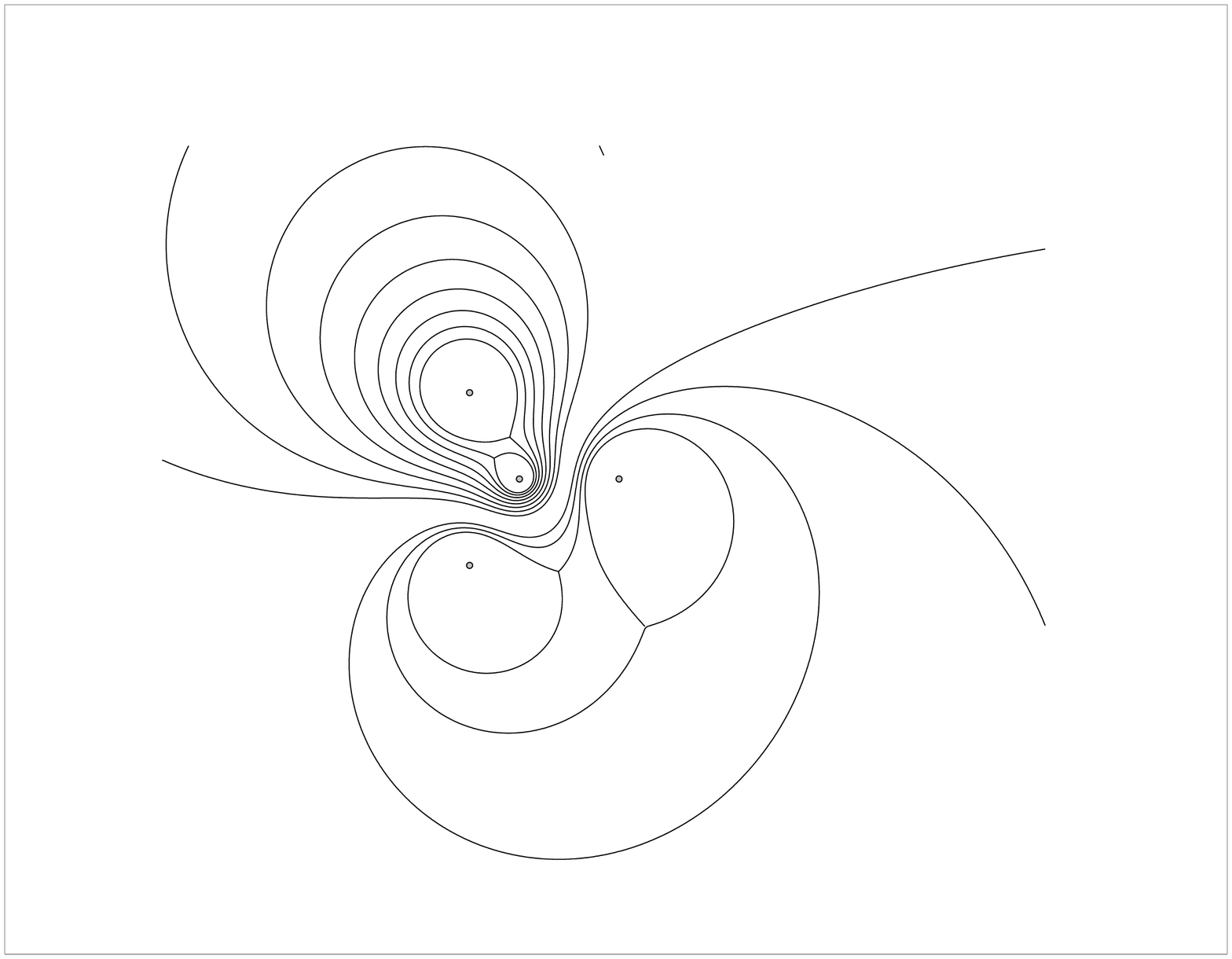,width=2.6in}}
    \put(4.3,3.75){$\Im{\rm m}\,a\neq0$} \put(4.3,3.55){$\Im{\rm
    m}\,b\neq0$} \put(2.7,3.75){case (4)}
    \put(0,0){\psfig{figure=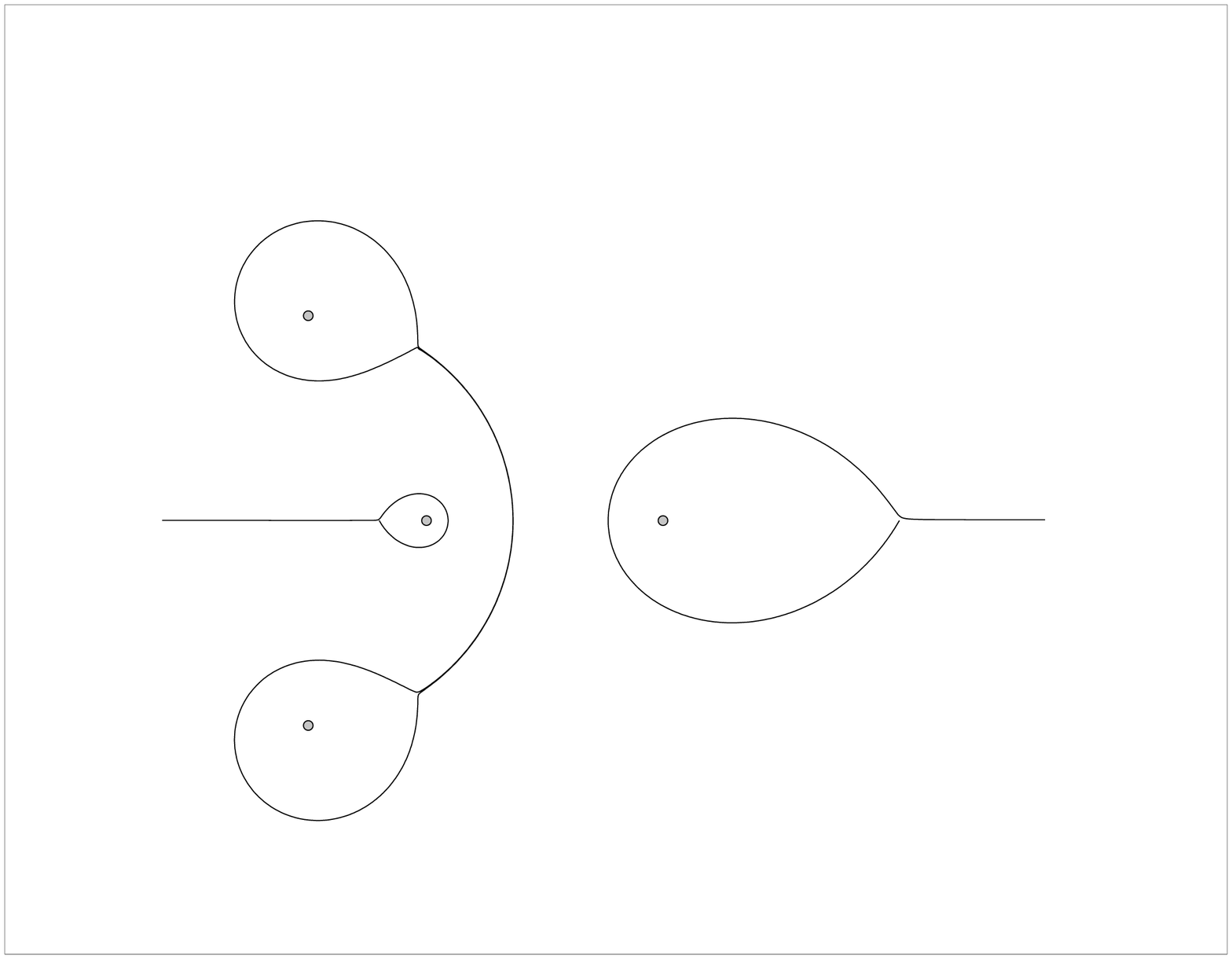,width=2.6in}}
    \put(1.8,1.75){$a>2\pi$} \put(0.1,1.75){case (2)}
    \put(1.7,1.55){$\Im{\rm m}\,b\neq0$}
    \put(2.6,0){\psfig{figure=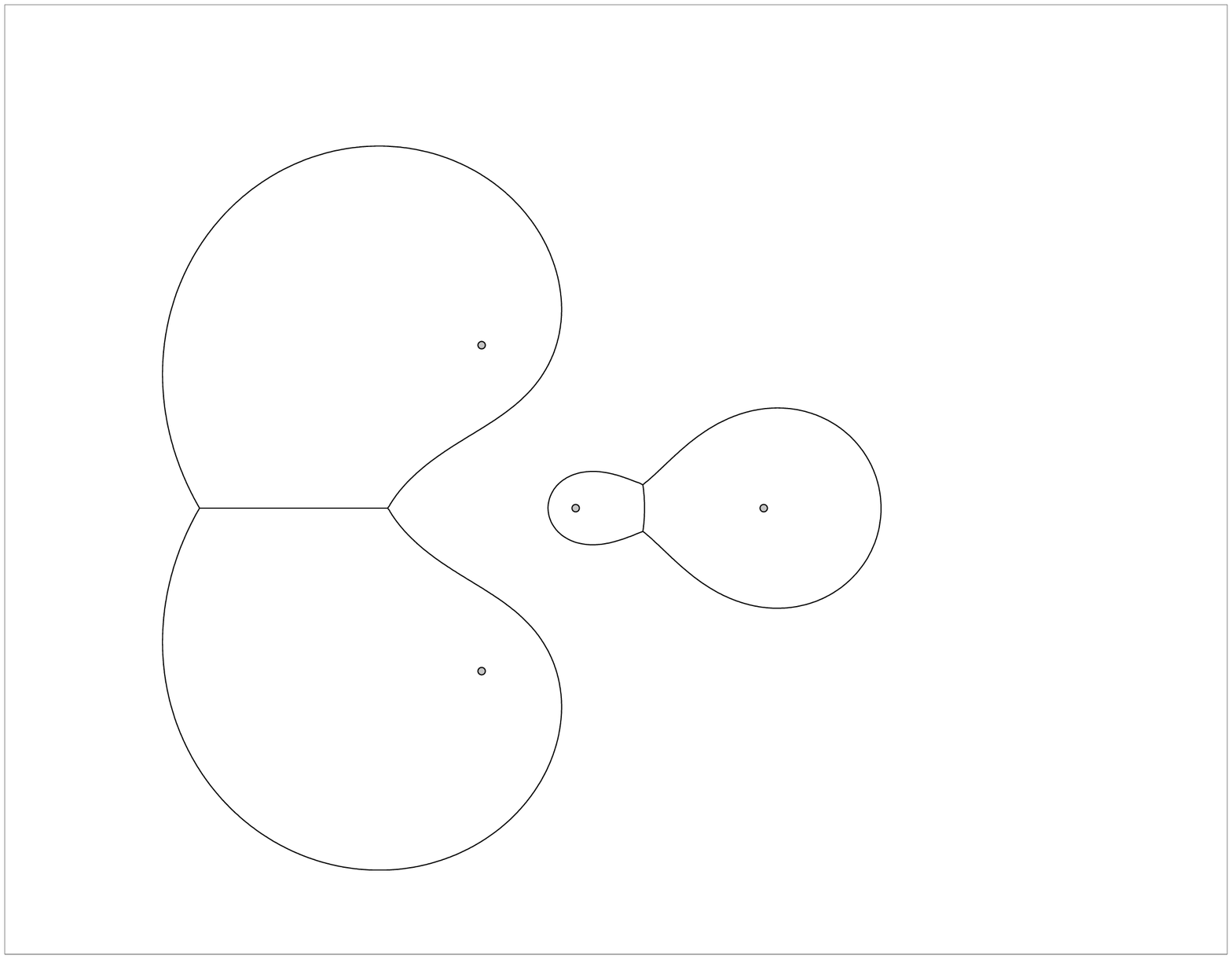,width=2.6in}}
    \put(4.4,1.75){$a<2\pi$} \put(4.3,1.55){$\Im{\rm m}\,b\neq0$}
    \put(2.7,1.75){case (3)} \put(3.25,1.0){\tiny $a$} \put(3.2,.3){\tiny
    $2\pi-a$} \put(3.2,1.5){\tiny $2\pi-a$}
  \end{picture}
\end{center}
\caption{Four different kinds of the critical graph.}
  \label{fig:graph}
\end{figure}

Any meromorphic function with two periods (an elliptic function) can be
written in terms of two basic elliptic functions --- the Weierstrass
$\wp$-function and its derivative $\wp'$ (see Ref.~\cite{lang}).  Let $u_0$
be the position of a pole which is inside the parallelogram $[\o_1, \o_2]$.
An elliptic function having two poles with residue $\pm i$ and a double
zero is uniquely defined by the positions of the zero an one of the
poles. Let the zero be at $u=0$ (we can always shift $u$ by a constant in
order to achieve this), and the pole with residue $i$ be at $u=u_0$, then
\begin{equation}
  \begin{array}{rcl}
    f(u)&=&\ds{i\wp'(u_0)\over \wp(u) - \wp(u_0)}\\[12pt] &=&\ds
              i(\z(u+u_0) - \z(u-u_0) - 2\z(u_0))\\[12pt] &=&\ds
              i{\d\over\d u} \left( \ln
              \frac{\sigma(u+u_0)}{\sigma(u-u_0)}-2\z(u_0)u \right),
  \end{array}
  \label{f(u)}
\end{equation}
where $\wp$, $\z$ and $\sigma$ are the corresponding Weierstrass functions
for the lattice $[\o_1,\;\o_2]$.  Using (\ref{abc:def}) and (\ref{f(u)}) we
can calculate $a$ and $b$:
\begin{equation}
  \begin{array}{rcccl}
    a&=&\ds-\int_0^{\o_1}f(u)\;\d u&=& -2\,\pi-2\,i\,\lf(\z(u_0)\,\o_1 -
        \n_1\,u_0\rt),\\[.5cm] b&=&\ds\int_0^{\o_2}f(u)\;\d u&=&
        -2\,\pi+2\,i\,\lf(\z(u_0)\,\o_2 - \n_2\,u_0\rt).
      \end{array}
  \label{abc}
\end{equation}
See Fig.~\ref{fig:sphere} to justify the limits of integration.  The values
of $a$ and $b$ define the geometry of the critical horizontal
trajectories. Using the last equation in (\ref{f(u)}) we can write the
quadratic differential as $\phi=(\d v)^2$, where
\begin{equation}
  v(u)=i\,\ln\frac{\sigma(u_0+u)}{\sigma(u_0-u)}-2\,i\, \z(u_0)\,u.
  \label{v}
\end{equation}
On the $v$ plane horizontal trajectories are horizontal lines.  From
(\ref{v}) and (\ref{abc}) we can see that on the $v$-plane the zeros of
$\phi$ are at $v(0)=0$, $v(\o_1)=a$, $v(\o_2)=b$.  Thus when $a$ and $b$
are real any three of the zeros are connected by one horizontal trajectory
and the critical graph is a tetrahedron. If only $a$ is real the critical
horizontal trajectories form two separate connected graphs.  When
$a\leq2\,\pi$ the zeros are connected in two pairs, each pair having three
horizontal trajectories traversing from one zero to the other. When
$a\geq2\,\pi$ we have a different picture, with each pair of zeros having
one trajectory passing between them and the others forming two
tadpoles. Finally, when none of the $a$, $b$ or $c$ is real, two of the
three critical trajectories leaving a zero collide on their way around a
pole and come back forming a tadpole and the other becomes infinite. Figure
\ref{fig:graph} illustrates these four cases.

\section{Four-string vertex and Feynman region}
\label{Miss}
\setcounter{equation}{0} In this section we show how integral invariants
can be used to find the four-string vertex. The use of complex values of
the integral invariants will allow us to describe the quadratic
differentials used to define local coordinates in the string vertex and
Feynman regions similarly using particular constraints imposed on the
possible values of the invariants.
\begin{figure}[ht]
  \begin{center}
    \leavevmode \setlength{\unitlength}{1.in}
  \begin{picture}(5.2,2)
    \put(0,0){\psfig{figure=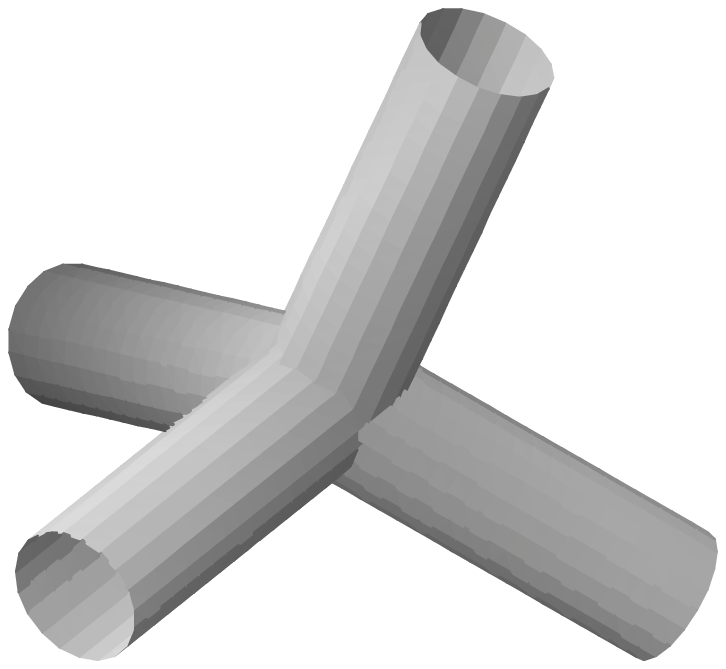,width=2.6in}} \put(0.3,0){(a)
    Strebel} \put(2.6,0){\psfig{figure=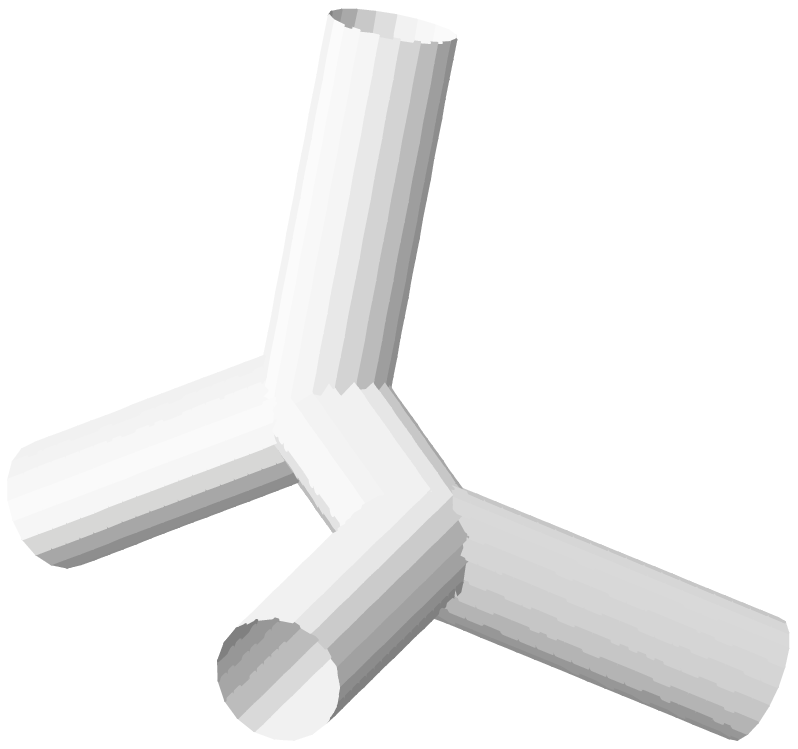,width=2.6in}}
    \put(2.9,0){(b) Feynman}
  \end{picture}
  \end{center}
  \caption{Riemann surfaces corresponding to a Strebel (a) and
    Feynman (b) quadratic differentials}
  \label{fig:feyn-diag}
\end{figure}

As was shown in Ref.~\cite{poly} the elementary interaction can be found
using so-called `Strebel quadratic differentials'. A Strebel quadratic
differential is a quadratic differential whose critical graph is a
polyhedron, or, --- as the analysis in sect.~\ref{Quad} shows --- all the
integral invariants are real. For the case of four-punctured spheres we
define the {\em Strebel constraint} by
\begin{equation}
  \label{st-cons}
  \im\,a=\im\,b=\im\,c=0.
\end{equation}
Given a quadratic differential $\phi=\varphi(z)(\d z)^2$ one can naturally
define a metric by $g = |\varphi(z)|\d z\d\bar z$. Since $\varphi(z)$ is
meromorphic, this metric has zero curvature at every point where
$\varphi(z)\neq0$:
\begin{equation}
  \label{curvature}
  R = -{4\over |\varphi(z)|^2} {\partial\over\partial z}
 {\partial\over\partial\bar z}\log |\varphi (z)| = 0.
\end{equation}
Therefore if we cut the sphere along the critical graph it will break into
pieces each isometric to a cylinder. For the Strebel quadratic differential
the four-punctured sphere breaks into four semi-infinite cylinders each of
circumference $2\pi$ (Fig.~\ref{fig:feyn-diag}). In order to reconstruct
the Riemann surface one has to glue these four cylinders along the edges of
a tetrahedron with the sides equal $a$, $b$ and $c$ (see
Ref.~\cite{zwie-cov}).

\begin{figure}[t]
  \begin{center}
    \leavevmode \psfig{figure=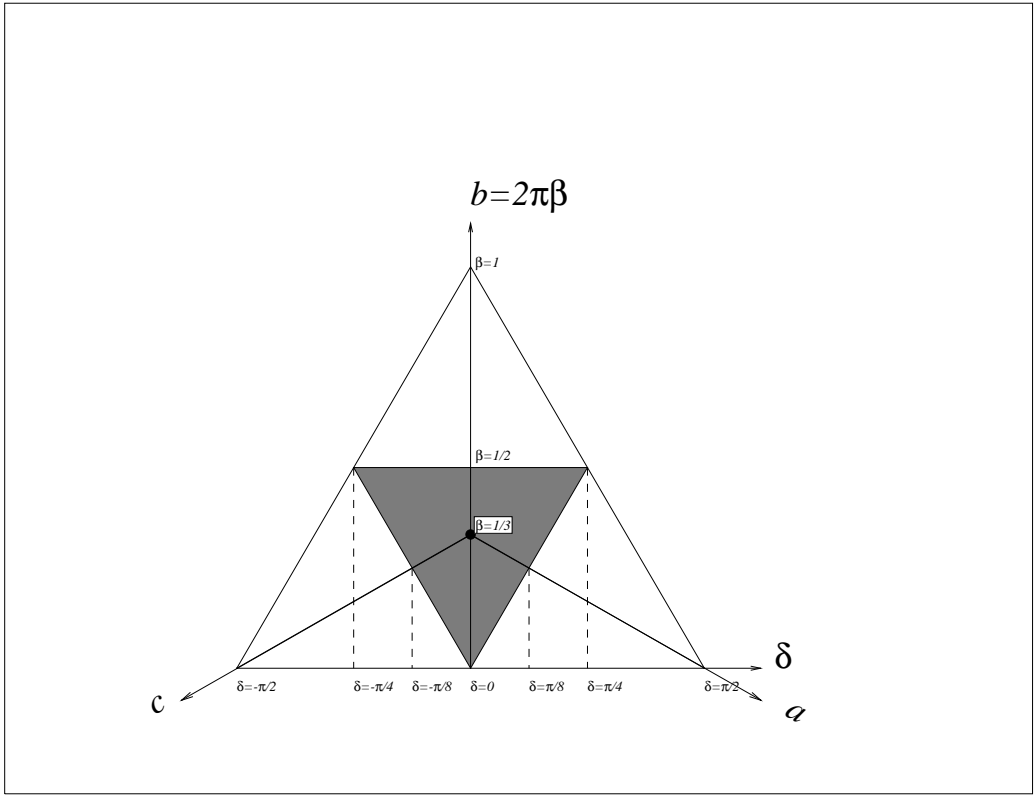,width=\the\textwidth}
  \end{center}
  \caption{The four-string vertex $\V_{0,4}$ on the $a+b+c=2\pi$ plane.}
  \label{fig:abc}
\end{figure}

Due to the Strebel theorem \cite{strebel} one can use real positive values
of the integral parameters ($a+b+c=2\pi$) in order to parameterize $\m$. It
is well known that we actually need two copies of the $abc$ triangle
$a+b+c=2\pi$ to cover the whole whole $\m$. This parameterization is very
useful because we can easily describe the four-string vertex $\V_{0,4}$
which is given by (see Ref.~\cite{poly})
\begin{equation}
a>\pi,\quad b>\pi\quad \mbox{and}\quad c>\pi.
  \label{V4def}
\end{equation}
In Fig.~\ref{fig:abc} we present the view at $abc$ triangle along the line
$a=b=c$. The shaded region corresponds to $\V_{0,4}$.

In order to calculate the contribution of Feynman diagrams we have to
define the measure $\mu$ in the Feynman region of the moduli space. We will
achieve this goal by finding the corresponding quadratic differential for
each Riemann surface or a string diagram in the Feynman region.

A Feynman string diagram for the four-string scattering is a Riemann
surface obtained by gluing together five cylinders with circumference
$2\pi$: four semi-infinite cylinders representing the scattering strings
and one finite cylinder representing an intermediate string or a
propagator. There are three topologically inequivalent ways to glue these
cylinders together corresponding to three channels $s$, $t$ and $u$.
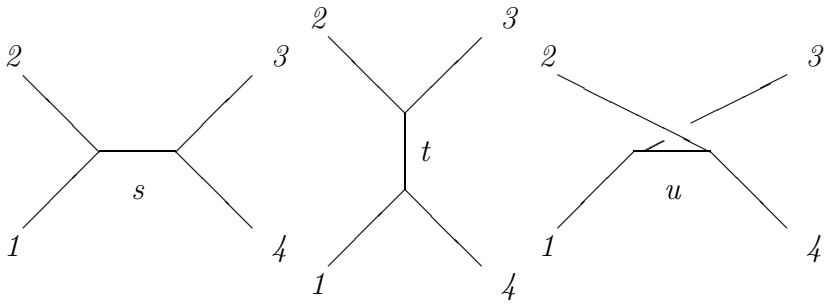
\begin{figure}[h]
  \begin{center}
    \leavevmode
\setlength{\unitlength}{0.01in}%
\begingroup\makeatletter\ifx\SetFigFont\undefined
% extract first six characters in \fmtname
\def\x#1#2#3#4#5#6#7\relax{\def\x{#1#2#3#4#5#6}}%
\expandafter\x\fmtname xxxxxx\relax \def\y{splain}%
\ifx\x\y   % LaTeX or SliTeX?
\gdef\SetFigFont#1#2#3{%
  \ifnum #1<17\tiny\else \ifnum #1<20\small\else \ifnum
  #1<24\normalsize\else \ifnum #1<29\large\else \ifnum #1<34\Large\else
  \ifnum #1<41\LARGE\else \huge\fi\fi\fi\fi\fi\fi
  \csname #3\endcsname}%
\else \gdef\SetFigFont#1#2#3{\begingroup \count@#1\relax \ifnum
25<\count@\count@25\fi
  \def\x{\endgroup\@setsize\SetFigFont{#2pt}}%
  \expandafter\x \csname \romannumeral\the\count@ pt\expandafter\endcsname
    \csname @\romannumeral\the\count@ pt\endcsname
  \csname #3\endcsname}%
\fi \fi\endgroup
\begin{picture}(420,155)(70,665)
\thinlines \put(120,740){\line(-1, 1){ 40}} \put( 80,700){\line( 1, 1){
40}} \put(120,740){\line( 1, 0){ 40}} \put(160,740){\line( 1, 1){ 40}}
\put(160,740){\line( 1,-1){ 40}} \put(280,760){\line( 0,-1){ 40}}
\put(400,740){\line( 1, 0){ 40}} \put(400,740){\line(-1,-1){ 40}}
\put(440,740){\line( 1,-1){ 40}} \put(360,780){\line( 2,-1){ 80}}
\multiput(400,740)(0.50000,0.25000){21}{
\makebox(0.1111,0.7778){\SetFigFont{5}{6}{rm}.}}
\put(430,755){\line( 2, 1){ 50}} \put(240,680){\line( 1, 1){ 40}}
\put(280,720){\line( 1,-1){ 40}} \put(280,760){\line( 1, 1){ 40}}
\put(280,760){\line(-1, 1){ 40}} \put(
70,685){\makebox(0,0)[lb]{\smash{\SetFigFont{12}{14.4}{it}1}}} \put(
70,785){\makebox(0,0)[lb]{\smash{\SetFigFont{12}{14.4}{it}2}}}
\put(210,785){\makebox(0,0)[lb]{\smash{\SetFigFont{12}{14.4}{it}3}}}
\put(210,685){\makebox(0,0)[lb]{\smash{\SetFigFont{12}{14.4}{it}4}}}
\put(230,665){\makebox(0,0)[lb]{\smash{\SetFigFont{12}{14.4}{it}1}}}
\put(230,805){\makebox(0,0)[lb]{\smash{\SetFigFont{12}{14.4}{it}2}}}
\put(330,805){\makebox(0,0)[lb]{\smash{\SetFigFont{12}{14.4}{it}3}}}
\put(330,665){\makebox(0,0)[lb]{\smash{\SetFigFont{12}{14.4}{it}4}}}
\put(350,685){\makebox(0,0)[lb]{\smash{\SetFigFont{12}{14.4}{it}1}}}
\put(490,785){\makebox(0,0)[lb]{\smash{\SetFigFont{12}{14.4}{it}3}}}
\put(490,685){\makebox(0,0)[lb]{\smash{\SetFigFont{12}{14.4}{it}4}}}
\put(350,785){\makebox(0,0)[lb]{\smash{\SetFigFont{12}{14.4}{it}2}}}
\put(140,715){\makebox(0,0)[b]{\smash{\SetFigFont{12}{14.4}{it}s}}}
\put(290,735){\makebox(0,0)[b]{\smash{\SetFigFont{12}{14.4}{it}t}}}
\put(420,715){\makebox(0,0)[b]{\smash{\SetFigFont{12}{14.4}{it}u}}}
\end{picture}
  \end{center}
  \caption{The three Feynman diagrams built with the three-vertex and
    propagator that enter in the computation of the four tachyon
    amplitude. }
  \label{fig:feyn}
\end{figure}
For each channel we can vary the length of the propagator $l$ and the twist
angle $\theta$.  This construction defines three non-intersecting regions
in the moduli space $\F_s$, $\F_t$ and $\F_u$ each naturally parameterized
by $l>0$ and $0<\theta<2\pi$. A Feynman string diagram can be easily
constructed using a quadratic differential with complex integral
invariants. Take a look at the case~$4$ in Fig.~\ref{fig:graph}, which
shows the critical graph of a quadratic differential which has one of the
integral invariants ($a$) real and less then $2\pi$. The correspondent
Riemann surface consists of two pairs of semi-infinite cylinders glued to a
finite cylinder with length $|\im\,b|$ and circumference $4\pi - 2 a$. If
we define the twist $\theta$ as an angle between two zeros on the
propagator we obtain $\theta = \re\,b$. Thus we conclude that in order to
define a a Feynman string diagram a quadratic differential should have one
integral invariant equal to $\pi$ and another equal to $\theta + il$.  We
define three {\em Feynman constraints} corresponding to the diagrams in
Fig.~\ref{fig:feyn} by
\begin{equation}
  \begin{array}{l}
    F_s:\quad a = \pi, \quad c = \th+i\,l;\cr F_t:\quad c = \pi, \quad b =
    \th+i\,l;\cr F_u:\quad b = \pi, \quad a = \th+i\,l.\cr
  \end{array}
  \label{feynstu}
\end{equation}
By definition the length of the propagator $l>0$ and the twist $\th$ is
between zero and $2\pi$. It is convenient to combine $l$ and $\th$ into one
complex variable $\e=e^{i\th-l}$ (for different channels $\e$ is equal to
either $e^{ia}$ or to $e^{ib}$ or to $e^{ic}$). Different values of $\e$
correspond to different Riemann surfaces or different points in
$\m$. Therefore each Feynman constraint defines a section over the
correspondent region in the moduli space. We define three regions $\F_s$,
$\F_t$ and $\F_u$ as the projections of the correspondent sections on
$\m$. Each of these regions can be naturally parameterized by $|\e|<1$. We
can summarize this construction on the following diagram
\begin{equation}
  \label{const-reg}
  \begin{array}{rcl}
    U=\{|\e|<1\}&\stackrel{F_{s,t.u}}{\longrightarrow}&\D^{R}\cr & &
    \downarrow\cr & & \F_{s,t,u}\subset\m.
  \end{array}
\end{equation}
 We also obtain an alternative description of the four-string vertex:
$\V_{0,4} = \m\backslash (\F_s\cup\F_t\cup\F_u)$. One can easily see that
this agrees with (\ref{V4def}).

Both the Feynman and the Strebel constraints define two-dimensional
subspaces in the four-dimensional $\D^R$, but these subspaces are quite
different. The Strebel constraint defines a global section of $\D^R$ over
$\m$. This is a result known as the Strebel theorem \cite{strebel}. The
section defined by the Strebel constraint is not holomorphic because the
constraint is given in terms of real functions on $\D^R$
(\ref{st-cons}). The Feynman constraints are defined by fixing a value of
one of the three holomorphic functions on $\D^R$: $a=\pi$, $b=\pi$ or
$c=\pi$. It is well known that the Feynman constraints define holomorphic
sections only over a part of $\m$, namely over the Feynman regions
$\F_{s,t,u}$.

Using complex integral invariants allows us to treat the four-string vertex
and the Feynman regions in a unified manner by imposing some extra
conditions (\ref{st-cons}) and (\ref{V4def}) or (\ref{feynstu}) on $a$, $b$
and $c$ and integrating over simple regions which they define.

At this point we face a dilemma: the measure of integration in the formulae
defining the four-tachyon amplitude (\ref{gst}) is given in terms of
$\chi$-invariants. On the other hand, the regions of integration for in the
definition of the elementary four-tachyon coupling and the formula defining
the massive states correction are given in terms of $a$, $b$ and $c$.
Therefore, our next goal will be to relate the $\chi$-invariants and $a$,
$b$ and $c$.  We will proceed in two steps: in the sect.~\ref{Main} we will
solve the system (\ref{abc}) and find the torus modulus $\tau=\o_1/\o_2$
and and the position of the pole $u_0$ in terms of $a$ and $b$.  Then, in
sect.~\ref{Infi}, we will express the $\chi$ invariants in terms of $\tau$
and $u_0$.

\section{The main equation}
\label{Main}
\setcounter{equation}{0} In this section we will explore the
system~(\ref{abc}). Let us fix the scale of the coordinate on the torus so
that $\o_1=\tau$ and $\o_2=1$, then the system~(\ref{abc}) can be written
as
\begin{equation}
\lf\{\begin{array}{l}
    \ds\a=1+{i\over\pi}\Big(\z(u_0;\,\tau)\,\tau-\n_1(\tau)\,u_0\Big)
    \\[12pt]
    \ds\b=1-{i\over\pi}\Big(\z(u_0;\,\tau)-\n_2(\tau)\,u_0\Big)
  \end{array}\rt.
  \label{albet}
\end{equation}
where
\begin{equation}
  \label{albedef}
   \a={a\over2\pi}\qquad \mbox{and}\qquad\b={b\over2\pi}.
\end{equation}
This is a system of two equations for two complex variables $\tau$ and
$u_0$, and its solution should define $\tau(\a,\b)$ and $u_0(\a,\b)$.  In
the present form it is extremely hard to solve. Fortunately we can reduce
this system to a single equation defining $\tau(\a,\b)$.  Using the
Legendre relation $\n_2(\tau)\tau-\n_1(\tau)=2\pi i$ we can deduce that the
system (\ref{albet}) is equivalent to
\begin{equation}
\lf\{ \begin{array}{rcl} u_0 &=&\ds{1-\b\over2}\tau+{1-\a\over2},\\[12pt]
    \z(u_0)&=&\ds{1-\b\over2}\n_1(\tau)+{1-\a\over2}\n_2(\tau).\\[12pt]
  \end{array}\rt.
  \label{legendre}
\end{equation}
Now we can eliminate $u_0$ and get
\begin{equation}
  \z\lf({1-\b\over2}\,\tau+{1-\a\over2}\,\mbox{\Large{;}}\;\tau\rt)=
  {1-\b\over2}\,\n_1(\tau)+{1-\a\over2}\,\n_2(\tau).
  \label{main1}
\end{equation}
This equation plays the major role in our approach to the four-string
amplitude problem. If we knew its solution $\tau(\a,\b)$ we would know the
solution to the system (\ref{albet}) because $u_0(\a,\b)$ is given by:
\begin{equation}
  \label{u0tau}
    u_0(\a,\b)={1-\b\over2}\tau(\a,\b)+{1-\a\over2}
\end{equation}
We will refer to (\ref{main1}) as the {\em main equation\/}.

In this section we will discuss the symmetries of this equation and find
two regions for $\a$ and $\b$ which correspond to large values of
$\im\,\tau$. When $\im\,\tau$ is large the $\z$ function can be expanded as
a series with respect to a small parameter $\qt=\exp(2\pi i\tau)$. We will
call this series the $q$-series. We will use a truncated $q$-series to find
approximate solutions of the main equation.  Then we return to the Strebel
case of real $\a$ and $\b$ and investigate the map from the $abc$ to the
$\tau$ plane.

\noindent \underline{Symmetries.} Recall that $\a$ and $\b$ represent three
invariants $a$, $b$ and $c$ which satisfy $a+b+c=2\,\pi$.  A permutation of
$a$, $b$ and $c$ is equivalent to a permutation of the zeros. The torus
modulus $\tau$ is closely related to the cross ratio of the zeros, and
permutation of the zeros results in a modular transformation on the $\tau$
plane.  More specifically:
\begin{equation}
  \begin{array}{rcllrcllrcl}
    a&\exch&b &\equiv&\a&\exch&\b &\equiv& \tau&\to& -1/\tau\cr b&\exch&c
    &\equiv&\b&\to&1-\a-\b&\equiv& \tau&\to& (2-\tau)/(1-\tau)\cr c&\exch&a
    &\equiv&\a&\to&1-\a-\b&\equiv& \tau&\to& -(1-\tau)/(2-\tau)\cr
  \end{array}
  \label{symms}
\end{equation}
One can easily check that the transformations (\ref{symms}) do not violate
(\ref{main1}) using modular properties of the $\z$-function.

Using the addition theorem for the $\z$ function (see
Ref.~\cite{functions}) one can show that the change of $\a$ and $\b$ to
$-\a$ and $-\b$ does not change eqn.~(\ref{main1}). This is quite obvious
because the integral invariants are defined up to a common sign which comes
from the ambiguity in taking a square root.

\noindent \underline{$\b\to0$ limit.}  Let us rewrite the second equation
of the system (\ref{albet}) using a $q$ expansion for the Weierstrass
$\z$-function (see Ref.~\cite[page 248]{lang})
\begin{equation}
  \label{zqexp}
 \z(u)=\n_2 u + \pi i{q_u+1\over q_u - 1} + 2\pi i \sum_{n=1}^\infty\lf[
           {\qt^n\quO\over1-\qt^n/\quO} - {\qt^n\,\quO\over1-\qt^n\,\quO}
           \rt],
\end{equation}
where we use a notation $q_x=\exp(2\pi i x)$. Terms linear in $u$ in the
expression for $\b$ cancel and we get
\begin{equation}
  {\b\over2}=\sum_{n=0}^\infty\lf[ {\qt^n(\qt/\quO)\over1-\qt^n(\qt/\quO)}
                       - {\qt^n\,\quO\over1-\qt^n\,\quO} \rt].
  \label{qmain}
\end{equation}
The reason why we collected the terms $\qt/\quO$ will be clear in a
moment. Exponentiating the first equation in (\ref{legendre}) we can
express $\quO$ in terms of $\qt$, $\a$ and $\b$ as
\begin{equation}
  \quO = -\qt^{\half}e^{-\pi i (\a +\b\tau)}
  \label{qu}
\end{equation}
If we substitute the value of $\quO$ from eqn.~(\ref{qu}) in to
eqn.~(\ref{qmain}) we will get an equation which is equivalent to the main
equation. Analyzing equations (\ref{qmain}) and (\ref{qu}) we conclude that
in the limit $\b\to0$
$$ \qt\semi\b^2\qquad \mbox{and} \qquad \quO\semi\b. $$ Therefore, in this
limit $\quO\semi\qt/\quO$ which is reflected in the way we wrote
(\ref{qmain}).  Moreover, $\qt$ being small in this limit allows us to find
an approximate solution to the main equation.  The first two terms in
$\b$-expansion of $\qt$ give
\begin{equation}
  \qt=-{\b^2\over16\,\cos^2\dl} -
  {i\,\sin\dl\over16\,\cos^3\dl}\lf(2\,\ln{4\,\cos\dl\over\b}-1\rt)\b^3
  +\O(\b^4),
  \label{qt3}
\end{equation}
where
\begin{equation}
  \dl=\pi\,\lf(\a+{\b-1\over2}\rt).
  \label{nots1}
\end{equation}
Taking the logarithm of (\ref{qt3}) we find
\begin{equation}
    \tau =\ds{1\over2}+{i\over\pi}\,\ln{4\,\cos\dl\over\b}
    +{\tan\dl\over2\pi}\lf(2\,\ln{4\,\cos\dl\over\b}-1\rt)\b +\O(\b^2).
\label{t2}
\end{equation}
This solution is valid for complex values of $\a$ and $\b$. Therefore it
can be used both for the vertex and the Feynman regions. The limit $\b\to0$
corresponds to the corner of the vertex (see~Fig.~\ref{fig:abc}) for real
$\b$ and to the limit of short propagator and small twist for $\im\,\b>0$.

\noindent \underline{$\im\,\a\to\infty$ limit.} There is another region
where $\im\,\tau(\a,\;\b)\to\infty$. This is the case when $0<\b<1$ is a
fixed real number and $\a\to i\,\infty$.  Indeed, from equation
(\ref{qmain}) we derive that in the limit $\qt\to0$ and finite $\b$
\begin{equation}
  \quO = -{\b\over2-\b},
  \label{qu0}
\end{equation}
so that $\quO$ is finite unless $\b=0$ or $\b=2$.  According to equation
(\ref{legendre})
\begin{equation}
 \a = 1 -2\,u_0 + (1-\b)\,\tau.
  \label{tu0}
\end{equation}
As we have seen, $u_0$ is finite as $\im\,\tau\to\infty$ and thus, for real
$\b$ and $\a\to\infty$
\begin{equation}
  \im\tau \semi {\a\over1-\b} .
  \label{tauab}
\end{equation}
Let us set $\b=1/2$, which corresponds to the Feynman ($u$-channel)
constraint. This constraint makes $\tau$ an analytic function of $\a$.  The
first equation of (\ref{legendre}) can now be written as
\begin{equation}
  u_0={1\over4}\,\tau+{1-\a\over2}, \quad \mbox{or}\quad
 \quO=-\qt^{1/4}\,q_\a^{-1/2},
  \label{u0}
\end{equation}
and collecting the terms of the same order, we can rewrite (\ref{qmain}) as
\begin{equation}
  {1\over4}=-{\quO\over1-\quO} + \sum_{n=1}^\infty\lf[
            {\qt^n/\quO\over1-\qt^n/\quO} - {\qt^n\,\quO\over1-\qt^n\,\quO}
            \rt].
  \label{qumain}
\end{equation}
Using (\ref{u0}) we can iterate (\ref{qumain}) and find $\qt$ as a power
series in $q_\a$.
\begin{equation}
  \qt= {{1\over {3^4}}}\,q_\a^2+ {{{512}\over {3^{10}}}}\,q_\a^4 +
       {{{94720}\over {3^{15}}}}\,q_\a^6 + {{{167118848}\over
       {3^{22}}}}\,q_\a^8+\O(q_\a^{10}),
  \label{feinq}
\end{equation}
or
\begin{equation}
  \begin{array}{rcl}
    \tau(\a,\;{1\over2})&=&\ds2\,\a +{2\,i\,\ln3\over\pi}\\[12pt]
    &&\ds-{i\over\pi}\lf( {256\over 3^6}q_\a^2+ {76544\over 3^{12}}q_\a^4+
    {99552256\over 3^{19}}q_\a^6+\O(q_\a^8)\rt).
  \end{array}
  \label{feint}
\end{equation}
The appearance of the powers of $3$ in the coefficients is quite
remarkable. Formula ( \ref{feint}) provides a good approximation for $\tau$
at large values of $\im\,\a$. It shows that in this limit $\tau$ is a
linear function of $\a$ with a finite intercept $2\,\ln3/\pi$. For small
$\im\,\a$ ( \ref{feint}) does not work, but we can still find an
approximate formula. All we have to do is exchange $\a$ and $\b$
in~(\ref{t2}). According to symmetry relations (\ref{symms}) this is
equivalent to $\tau\to -1/\tau$, therefore for $\b=1/2$ and small $\a$, we
have
\begin{equation}
  -{1\over\tau} = {i\over\pi}\ln{4\,i\over\a}+\O(\a^2).
  \label{smalla}
\end{equation}
\begin{figure}[ht]
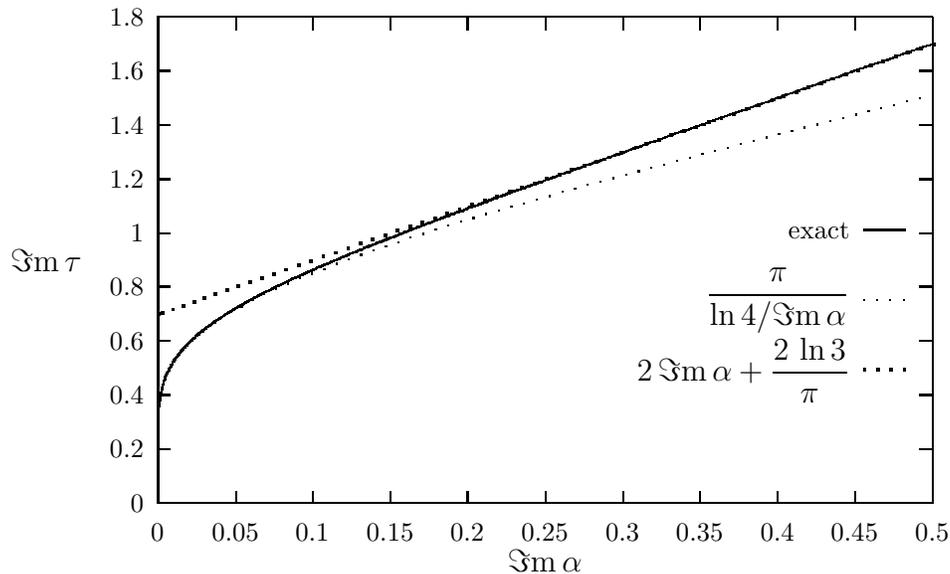

  \begin{center}
    \leavevmode
% GNUPLOT: LaTeX picture
\setlength{\unitlength}{0.240900pt}
\ifx\plotpoint\undefined\newsavebox{\plotpoint}\fi
\sbox{\plotpoint}{\rule[-0.200pt]{0.400pt}{0.400pt}}%


  \end{center}
  \caption{Solution of the main equation for $\b=1/2$ and imaginary
    $\a$.}
  \label{fig:tau-a}
\end{figure}
In Fig.~\ref{fig:tau-a} we show the result of numerical solution of the
main equation together with the first order approximations for small and
large $\im\,\a$.

\noindent \underline{The Strebel case.} We now return to the case when $\a$
and $\b$ are real and represent a point on the equilateral triangle
$a+b+c=2\pi$ where $a$, $b$ and $c$ are real and positive. Strebel's
theorem guarantees the existence of a solution to~(\ref{main1}) for every
point on the $abc$ triangle. Indeed, $\tau$ is related to $\lz$ by a
modular function of level 2, namely $\l(\tau)$ (see Ref.~\cite[page
254]{strebel} ).  This function maps it's fundamental domain $\Gamma$,
defined by
$$ \Gamma=\{\tau:\; -1<\re\,\tau\leq1,\; |2\,\tau - 1|\geq1\;
 \mbox{and}\;|2\,\tau + 1|>1\},
$$ bijectively to the whole complex plane. Therefore the existence of a
Strebel differential is equivalent to the existence of a solution to the
main equation in the fundamental domain of $\l(\tau)$.

Two Strebel differentials such that the zeros and poles of one are complex
conjugate to those of the other have the same set of $a$, $b$ and $c$
invariants. Therefore in the fundamental domain $\Gamma$ of $\l(\tau)$ we
should have two solutions to the main equation. These two solutions
correspond to conjugate values of $\l(\tau)$ and therefore are symmetric
with respect to the imaginary axis on the $\tau$ plane.  Finally, we
conclude that for every point inside the $abc$ triangle there exist a
solution to (\ref{main1}) satisfying $0\leq\tau\leq1$ and
$|2\,\tau-1|\geq1$. We will call this region $\Gamma/2$.

The main equation defines a map from the $abc$ plane to $\Gamma/2$.  Some
information about this map can be obtained from the symmetry.  According to
(\ref{symms}) the line $a=c$ ($\dl=0$) on the $abc$ plane maps on to the
line $\im\,\tau=1/2$. Similarly, $a=b$ maps on to the circle $|\tau|=1$ and
$b=c$ on to $|\tau-1|=1$, and we conclude that the most symmetric point
($a=b=c=2\pi/3$) is mapped to
\begin{equation}
  \tau\lf(\a={1\over3},\;\b={1\over3}\rt) = e^{\pi\,i\over3}.
  \label{sym}
\end{equation}

According to (\ref{t2}), the whole line $b=0$ maps on to the single point
$\tau=1/2+i\,\infty$, and therefore the other two sides of the $abc$
triangle $a=0$ and $c=0$ are correspondingly mapped to $0$ and $1$
respectively.  This seemingly leads to a contradiction at the corners. For
example when $a=b=0$ the solution must be $\tau=0$ because $a=0$; on the
other hand it should be $\tau=1/2+i\,\infty$ because $b=0$, but at the same
time it should be somewhere on the unit circle $|\tau|=1$ because $a=b$.
In fact there is no contradiction because if we rewrite the main equation
for this case we get
\begin{equation}
  \z\lf({\tau\over2}+{1\over2}\,\mbox{\Large{;}}\;\tau\rt)=
  {1\over2}\n_1(\tau)+{1\over2}\n_2(\tau),
  \label{main0}
\end{equation}
which is valid for {\em any} value of $\tau$.  The arbitrariness of $\tau$
does not contradict the Strebel theorem which guarantees the uniqueness of
the quadratic differential because as we will show in the next section, the
point $\a=\b=0$ corresponds to $\lp=0$ which is excluded from $\m$.  It is
interesting to investigate how the solution to (\ref{main1}) behaves in the
vicinity of a corner.

The corner $a=b=0$ corresponds to $\dl=-\pi/2$ (see Fig.~\ref{fig:abc}). It
is problematic to use the expression (\ref{qt3}) because the coefficients
diverge as $\dl\to-\pi/2$.

Let $\a$ and $\b$ be small, but not both equal to zero. Recall, that
$\z'(u)=-\wp(u)$ and $\wp(\tau/2+1/2)=e_3(\tau)$. When we keep only first
order terms in $\a$ and $\b$ in (\ref{main1}) we find
\begin{equation}
  -e_3(\tau)\lf({\a\over2}+{\b\over2}\tau\rt)
  ={\a\over2}\n_2(\tau)+{\b\over2}\n_1(\tau),
  \label{cor}
\end{equation}
then, using the Legendre relation to exclude $\n_1(\tau)$, we get
\begin{equation}
  (\n_2(\tau)+e_3(\tau))\lf({\a\over\b}+\tau\rt)=2\pi i.
  \label{cor1}
\end{equation}
Inspecting (\ref{cor1}) we, conclude that the limiting value of $\tau$
depends on the ratio $r=\b/\a$. Moreover for any value of $\tau$ there
exists an $r$ such that
$$ \lim_{\a\to0} \tau(\a,\;r\,\a) = \tau.
$$ From (\ref{cor1}) we can even find the ratio in terms of $\tau$:
\begin{equation}
  r=\lf[{2\pi i\over\n_2(\tau)+e_3(\tau)}-\tau\rt]^{-1}.
  \label{r(tau)}
\end{equation}
It is hard to tell what values of $\tau$ correspond to real $r$.  For large
$\im\,\tau$ we may use the $q$ expansion
\begin{equation}
  \n_2(\tau)+e_3(\tau)=8\pi^2\sum_{n=0}^\infty{\qt^{n+{1\over2}}\over
                                    1+\qt^{n+{1\over2}}},
  \label{ne}
\end{equation}
and solve (\ref{cor1}) approximately for $\tau(r)$
\begin{equation}
   \tau(r)={1\over2}+{i\over\pi}\ln{4\pi\over r}
   +\lf({i\over2}-{1\over\pi}\ln{4\pi\over r}\rt){r\over\pi} +{\rm O}(r^2).
  \label{kr}
\end{equation}
One can check that (\ref{qt3}) yields the same result in the limit of small
$\a$, $\b$ and $\b/\a$.
\begin{figure}[t]
  \begin{center}
    \leavevmode \psfig{figure=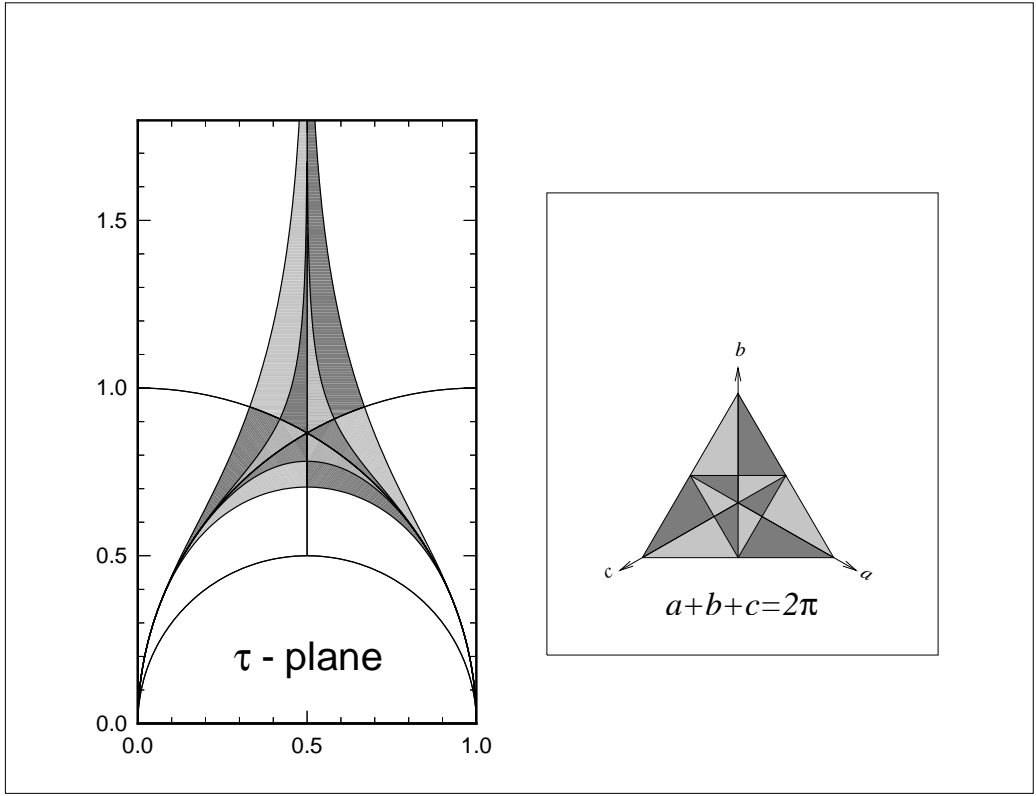,width=\the\textwidth}
  \end{center}
  \caption{Solution of the main equation for real $\a$ and $\b$.}
  \label{fig:tau}
\end{figure}
It is interesting that the map of the $abc$ triangle to the $\tau$ plane
does not cover $\Gamma/2$. It is mapped to a curved triangle.  The sides of
the original triangle ($a=0$, $b=0$ and $c=0$) become the corners
($\tau=0$, $\tau=\infty$ and $\tau=1$), while the corners blow up and
become sides. Fig.~\ref{fig:tau} represents a map from the $a+b+c=2\pi$
plane to the $\tau$-plane. The corresponding regions of the $\tau$ and
$abc$ planes are shaded with matching gray levels on the plot.

\section{Infinite products}
\label{Infi}
\setcounter{equation}{0} In this section we will perform the second step of
the program announced at the end of sect.~\ref{Miss}. We will derive
explicit formulae for the $\chi$ invariants as functions of the torus
modular parameter $\tau$ and the position of the pole $u_0$. We will find
$\lp=-\chi_s/\chi_u$ and $\lz$. The latter will be found as a special case
$u_0=0$ of a formula defining $\lp$.

Recall that the $\chi$ invariants are defined in terms of the positions of
the poles and the mapping radii as
\begin{equation}
\chi_{ij}={(z_i-z_j)^2\over\rho_i\rho_j}.
  \label{chiij1}
\end{equation}
As before, the coordinate $u$ on the torus is fixed by $\o_2=1$. We can
choose the coordinate $z$ on the sphere so that $z=\wp(u/2)$. The positions
of the poles on the sphere are given by
\begin{equation}
    z_i=\ds \wp\lf({u_i\over2}\rt), \quad u_i = u_0+\o_i,\qquad
i=1,\dots,4,
  \label{zi}
\end{equation}
where $\o_1=\tau$, $\o_2=1$, $\o_3=1+\tau$ and $\o_4=0$. So far, the only
nontrivial part of (\ref{chiij1}) are the mapping radii. Due to the
translational symmetry all four mapping radii of the coordinate disks on
the torus are equal and we denote their common value value by $\rho$.
According to the general procedure described in sect.~\ref{Howa}, in order
to calculate $\rho$ we have to find a local coordinate $w$ around $u_0$
such that locally
$$ \varphi=-{(\d w)^2\over w^2},\qquad\mbox{and}\qquad w(0)=1.
$$ The last condition fixes the scale of $w$ as well as its phase.  From
equation (\ref{f(u)}) we derive
\begin{equation}
  w(u) = {\s(u-u_0)\over\s(u+u_0)}\,e^{2\,\z(u_0)\,u}.
  \label{w}
\end{equation}
Note that $w(u)$ is just an exponent of the function $v(u)$ introduced in
sect.~\ref{Quad}, $w(u) = \exp(i\,v(u))$.  The mapping radius is the
inverse of the derivative of $w(u)$ at $u_0$.
\begin{equation}
  \rho^{-1}=w'(u_0)={e^{2\,\z(u_0)\,u_0}\over\s(2\,u_0)}.
  \label{w'}
\end{equation}
When we go from the torus to the sphere we make a change of coordinates
from $u$ to $z=\wp(u/2)$, therefore each mapping radius picks up a factor
of $(\d/\d u)\wp(u/2)$ and we find
\begin{equation}
  \rho_i = \half\wp'\lf(u_i\over2\rt)\rho.
  \label{rhoi}
\end{equation}
Now we can combine (\ref{zi}), (\ref{w'}) and (\ref{rhoi}) with
(\ref{chiij1}) to obtain
\begin{equation}
  \chi_{ij}=4\,{\lf(\wp\lf(u_i\over2\rt)-\wp\lf(u_j\over2\rt)\rt)^2
                      \over\wp'\lf(u_i\over2\rt)\wp'\lf(u_j\over2\rt)}
                      \,\rho^{-2}.
  \label{chip}
\end{equation}
This expression can be rewritten in terms of Weierstrass $\s$-functions.
We need the formulae for the difference of two $\wp$ functions (see
Ref.~\cite[page 243]{lang})
\begin{equation}
\wp(u)-\wp(v)=-{\s(u+v)\,\s(u-v)\over\s^2(u)\,\s^2(v)},
  \label{p-s}
\end{equation}
and their derivatives
\begin{equation}
\wp'(u)=-{\s(2\,u)\over\s^4(u)}.
  \label{wp's}
\end{equation}
The latter formula is just the derivative of (\ref{p-s}) with respect to
$v$ at the point $v=u$. Now we see that the powers of $\s(u_i/2)$ cancel
and we get
\begin{equation}
\chi_{ij}=4{\s^2\lf(u_0+{\o_i+\o_j\over2}\rt)
                       \s^2\lf({\o_i-\o_j\over2}\rt)\over
                       \s(u_0+\o_i)\s(u_0+\o_j)}\,\rho^{-2}.
  \label{chis}
\end{equation}
The prefactor of $\rho^{-2}$ in this expression is an elliptic function of
$u_0$ with periods $\o_1$ and $\o_2$, which was not obvious from
eqn. (\ref{chip}) because it was written in terms of elliptic functions of
$u/2$. This extra periodicity enforces the symmetry relations
(\ref{recsym}). We can further simplify eqn. (\ref{chis}) by introducing a
new function $\varphi(u)$ which is closely related to the Weierstrass
$\s(u)$ (see Ref.~\cite[page 246]{lang}),
\begin{equation}
\vp(u) = e^{-{1\over2}\n_2u^2}q_u^{1\over2}\s(u),
  \label{phis}
\end{equation}
where $q_u=e^{2\pi i u}$ and $\n_2=\half\z(\half)$ is a quasi-period of the
$\z$-function. This function has the following properties
\begin{equation}
  \varphi(u+1)=\varphi(u), \qquad\mbox{and}\qquad \varphi(u+\tau)=-{1\over
  q_u}\varphi(u).
  \label{phipro}
\end{equation}
We can use $\vp$ to replace $\s$ in eqn. (\ref{chis}) and we find
\begin{equation}
  \begin{array}{rcccccrl}
        \chi_s&=&\chi_{12}&=&\chi_{34}&=&- \ds4\,{\quO\over\qt^\half}&\ds
                   {\vp^2\lf(u_0+{\tau+1\over2}\rt)
                   \vp^2\lf({\tau+1\over2}\rt)\over
                   \vp^2(u_0)}\rho^{-2},\\[12pt]
                   \chi_t&=&\chi_{14}&=&\chi_{23}&=&
                   \ds~4\,{\quO\over\qt^\half}&\ds
                   {\vp^2\lf(u_0+{\tau\over2}\rt)
                   \vp^2\lf(\tau\over2\rt)\over
                   \vp^2(u_0)}\rho^{-2},\\[12pt]
                   \chi_u&=&\chi_{13}&=&\chi_{24}&=& 4&\ds
                   {\vp^2\lf(u_0+\half\rt) \vp^2\lf(\half\rt)\over
                   \vp^2(u_0)}\rho^{-2}.\\[12pt]
  \end{array}
  \label{chistuphi}
\end{equation}
The cross ratio of the poles does not depend on $\rho$ and we can find it
as
\begin{equation}
\lp(u_0) =- {\chi_s\over\chi_u} ={\quO\over\qt^\half}
                            {\vp^2\lf(u_0+{\tau+1\over2}\rt)
                            \vp^2\lf({\tau+1\over2}\rt)\over
                            \vp^2\lf(u_0+{1\over2}\rt)
                            \vp^2\lf({1\over2}\rt)}.
  \label{lpo}
\end{equation}

In the special case, $u_0=0$, this gives the cross ratio of the zeros
\begin{equation}
\lz = \lp(0) = \qt^{-\half}{\vp^4\lf({\tau+1\over2}\rt)\over
                             \vp^4\lf({1\over2}\rt)}.
  \label{lzo}
\end{equation}

The $\varphi$-function has a simple infinite product expansion in terms of
$\qu$ and $\qt$ (see Ref.~\cite[page 247]{lang}):
\begin{equation}
  \varphi(u;\;\tau)=(2\pi i)^{-1} (\qu - 1)\; \prod_{n=1}^\infty{(1 -
  \qt^n\,\qu)(1 - \qt^n/\qu)\over (1 - \qt^n)^2}.
  \label{phi-prod}
\end{equation}
This product converges as a power series with ratio $\qt$ for small values
of $\qt$. Note that by symmetry we can always choose $\tau$ to lie in the
fundamental region defined by $\lf|\re\,\tau\rt|\leq1/2$ and
$|\tau|\geq1$. The maximum value of $|\qt|$ in this region is obtained at
$\tau=(\pm1+i\sqrt{3})/2$, therefore
$$ |\qt|\leq\exp(-\pi\sqrt{3})\approx 0.00433.
$$ Such a small value of $|\qt|$ makes the product (\ref{phi-prod}) very
useful for numerical calculations.

The formulae (\ref{lpo}) and (\ref{lzo}) together with (\ref{phi-prod})
provide the infinite products for $\lp$ and $\lz$. In order to find similar
products for $\chi$'s we have to express the mapping radius $\rho$ in terms
of the function $\vp$,
\begin{equation}
  \rho^{-1}={e^{2u_0\z(u_0)}\over e^{2\n_2u_0^2}\quO^{-1}\vp(2u_0)}= {\quO
   e^{2u_0(\z(u_0)-\n_2u_0)}\over\vp(2u_0)}= {\quO^{{\b}}\over\vp(2u_0)},
  \label{btrick}
\end{equation}
where we use the second equation of the system (\ref{albet}) for $\b$.

For future reference we present here the products for all $\chi$'s.
\begin{equation}
  \begin{array}{rcl}
        \chi_s&=&\ds-4\,{\quO^{1+2\b}\over\qt^\half}
                      {1\over(1+\quO)^2(1-\quO)^4}\times\\[12pt] &
                      &\ds\quad\times\prod_{n=1}^\infty
                      {(1+\qt^{n-\half})^4\over
                      (1-\qt^n\quO^2)^2(1-\qt^n/\quO^2)^2}
                      {(1+\qt^{n-\half}\quO)^2(1+\qt^{n-\half}/\quO)^2\over
                      (1-\qt^n\quO)^2 (1-\qt^n/\quO)^2 },\\[12pt]
                      \chi_t&=&\ds~4\,{\quO^{1+2\b}\over\qt^\half}
                      {1\over(1+\quO)^2(1-\quO)^4}\times\\[12pt] &
                      &\ds\quad\times\prod_{n=1}^\infty
                      {(1-\qt^{n-\half})^4\over
                      (1-\qt^n\quO^2)^2(1-\qt^n/\quO^2)^2}
                      {(1-\qt^{n-\half}\quO)^2(1-\qt^{n-\half}/\quO)^2\over
                      (1-\qt^n\quO)^2 (1-\qt^n/\quO)^2 },\\[12pt]
                      \chi_u&=&\ds16\,{\quO^{2\b}\over(1-\quO)^4}\times\\[12pt]
                      & &\ds\quad\times\prod_{n=1}^\infty
                      {(1+\qt^{n})^4\over
                      (1-\qt^n\quO^2)^2(1-\qt^n/\quO^2)^2}
                      {(1+\qt^{n}\quO)^2(1+\qt^{n}/\quO)^2\over
                      (1-\qt^n\quO)^2 (1-\qt^n/\quO)^2 }.\\[12pt]
  \end{array}
  \label{chipro}
\end{equation}
It is not so easy to show that the sum of these products is zero as
required by eqn. (\ref{s+t+u}).

Dividing the first equation of (\ref{chipro}) by the third we find an
infinite product for the cross ratio of the poles:
\begin{equation}
\lp=-{\chi_s\over\chi_u} = {1\over4}{\quO\over(1+\quO)^2}
\prod_{n=1}^\infty{(1+\qt^{n-\half}\quO)^2(1+\qt^{n-\half}/\quO)^2\over
(1+\qt^{n}\quO )^2(1+\qt^{n}/\quO )^2}.
  \label{lampro}
\end{equation}

\section{From $a$, $b$ and $c$ to $\chi_s$, $\chi_t$ and $\chi_u$}
\label{From}
\setcounter{equation}{0} In this section we combine the results of the
previous two and investigate how $\chi$ invariants depend on $a$, $b$ and
$c$.

\noindent \underline{Exact results.}  There are very few cases when the
$\chi$ invariants can be found exactly. These are the cases when we know
the solution of the main equation. Such a solution is available for example
in the case of a degenerate quadratic differential i.e. when any of $a$,
$b$ or $c$ is zero.  According to eqn. (\ref{qt3}) $b=0$ corresponds to
$\qt=0$ and $\quO/\qt^{1/2}=-i\,e^{i\dl}$ and the $\chi$ invariants are
found to be
$$\chi_s=4\,{1-\sin\dl\over\cos\dl}e^{i\dl},\quad
  \chi_t=4\,{1+\sin\dl\over\cos\dl}e^{i\dl},\quad
  \chi_u=-\chi_s-\chi_t=-{8\over\cos\dl}e^{i\dl}$$ Note that for real $\dl$
  all the $\chi$ have the same phase, and therefore the cross ratio $\lp$
  is real:
\begin{equation}
\lp={1-\sin\dl\over2}
  \label{lpreal}
\end{equation}

The small parameter $\qt$ is also exactly zero in the limit
$\im\,\a\to\infty$ (see sect.~\ref{Main}). In this limit, $\quO$ is given
by (\ref{qu0}), and the $\chi$ invariants are
\begin{equation}
\chi_s=\infty,\quad\chi_t=\infty,\quad
\chi_u=-\chi_s-\chi_t=16\,{\quO^{2\b}\over(1-\quO)^4}
  \label{three}
\end{equation}
These results can also be obtained by an elementary approach. For example,
in the case $b=0$ and real $a$ and $c$, we can choose the uniformizing
coordinate $z$ so that the poles of the quadratic differential are located
at the vertices of a rectangle and the two degenerate zeros are at $0$ and
$\infty$. From symmetry, the horizontal trajectories are the symmetry lines
and we can find the mapping radii by making a conformal transformation.

The other case of infinite $\im\,\a$ corresponds to the degeneracy of the
poles. In this limit, two poles collide and we effectively have a three
punctured sphere. For the case $\b=\half$, this sphere is the Witten vertex
and the $\chi_u$ in the formula (\ref{three}) gives correct value
$|\chi|=3^3/2^4$.

The only nontrivial point where an exact solution is still available is the
most symmetric point $a=b=c$. In this case
$$ \tau=e^{i\pi\over3},
$$ which corresponds to the so-called equianharmonic case in the theory of
elliptic functions. In this case, all the necessary values of the
Weierstrass functions can be evaluated explicitly in terms of elementary
functions (see Ref.~\cite{functions}) and we obtain
\begin{equation}
\chi_s={2^5\sqrt[3]{2}\over3^2}e^{-{2\pi i\over3}},\quad
\chi_t={2^5\sqrt[3]{2}\over3^2}e^{2\pi i\over3},\quad
\chi_u={2^5\sqrt[3]{2}\over3^2}.
  \label{mostsym}
\end{equation}

The upper left picture on Fig.~\ref{fig:graph} shows the critical graph for
this case. It is formed by three straight lines connecting the first three
zeros with the last placed at infinity and three arcs connecting two finite
zeros having the center at the third.

\noindent \underline{Approximate results.} For other values of the integral
invariants no exact solution for the main equation is available, but we can
still solve perturbatively as we did in sect.~\ref{Main} and find
approximate formulae for the $\chi$'s.  Consider the case of large
$\im\,\a$ and $\b=\half$. In sect.~\ref{Main} we found the solution of the
main equation up to the $8$-th order of $q_\a$ (see eqn.(\ref{feinq})).
\begin{equation}
  \qt= {{1\over {3^4}}}\,q_\a^2+ {{{512}\over {3^{10}}}}\,q_\a^4 +
       {{{94720}\over {3^{15}}}}\,q_\a^6 + {{{167118848}\over
       {3^{22}}}}\,q_\a^8+\O(q_\a^{10}).
  \label{feinq1}
\end{equation}
We can find $\quO$ as
\begin{equation}
\quO = -{\qt^{1\over4}\over q_\a^\half}= - {1\over3} -
       {{128\over{3^7}}}\,q_\a^2 - {{{15488}\over {3^{12}}}}\,q_\a^4 -
       {{{7280128}\over {3^{18}}}}\,q_\a^6 + \O(q_\a^{8}).
  \label{feinqu}
\end{equation}
Using these values to substitute in (\ref{chistuphi}) we get the following
approximate formulae for $\chi$'s
\begin{equation}
  \begin{array}{rcl}
    \chi_s &=&\ds {{3^6}\over {2^8}}{1\over\qa} + {{3^3}\over {2^5}} +
                  {{3^2}\over {2^6}} \qa + {{5}\over {2\cdot3^3}} \qa^2 +
                  {{1609}\over {2^7\cdot3^6}}\qa^3\\[12pt]&&\ds\qquad +
                  {{343}\over {2\cdot3^8}}\qa^4 + {{16981}\over
                  {2^5\cdot3^{11}}}\qa^5 + {{163174}\over {3^{15}}}\qa^6 +
                  {{{\O}(\qa^7)}},\\[12pt] \chi_t &=&\ds - {{3^6}\over
                  {2^8}}{1\over\qa} + {{3^3}\over {2^5}} - {{3^2}\over
                  {2^6}} \qa + {{5}\over {2\cdot3^3}} \qa^2 - {{1609}\over
                  {2^7\cdot3^6}}\qa^3\\[12pt]&&\ds\qquad + {{343}\over
                  {2\cdot3^8}}\qa^4 - {{16981}\over {2^5\cdot3^{11}}}\qa^5
                  + {{163174}\over {3^{15}}}\qa^6 +
                  {{{\O}(\qa^7)}},\\[12pt] \chi_u &=&\ds - {3^3\over2^4} -
                  {{5\over{3^3}}}\,q_\a^2 - {{{343}\over {3^{8}}}}\,q_\a^4
                  - {{{326348}\over {3^{15}}}}\,q_\a^6 + \O(q_\a^{8}),
  \end{array}
  \label{chiqa}
\end{equation}
and the cross ratio
\begin{equation}
\l=-{\chi_s\over\chi_u}= {3^3\over 2^4}\,\qa^{-1} + {1\over 2} -
   {11\,\qa\over 2^2\cdot3^3} - {1621\over 2^3\cdot 3^8}\,\qa^3 -
   {413941\over 2\cdot3^{15}}\,\qa^5 + \O(\qa^7).
  \label{lamqa}
\end{equation}
As expected $\chi_s+\chi_t+\chi_u = 0$ up to this order.

For small $\b$ an approximate solution to the main equation is given by
(\ref{t2}). We can use this approximate solution together with the infinite
products (\ref{chipro}) and find $\chi$'s, but if we leave all the terms
the expressions become too complicated. We will need the full expression
depending on $a$ and $b$ only for the case of real $a$ and $b$. Most
interesting is the $\l$ dependence on $a$ and $b$, which describes the map
from $abc$ triangle to $\m$.  Keeping only the first non-vanishing terms in
both the real and imaginary part of $\lp$, we can write
\begin{equation}
  \lp={1-\sin\dl\over2}
     -i\,{\cos\dl\over2}\lf(1+\ln{4\,\cos\dl\over\b}\rt)\b+{\rm O}(\b^2),.
  \label{l2}
\end{equation}

\section{Summing the Feynman diagrams}
\label{Summ}
\setcounter{equation}{0} In this section we compute the part of the
tachyonic amplitude which comes from the Feynman diagrams. We show how to
express this partial amplitude in terms of an integral over a part of the
moduli space.  We analyze analytic properties of this integral and show
that it has no singularity at zero momentum. Our analysis allows us to
calculate the Feynman part of the amplitude at zero momentum.

We define the partial or Feynman amplitude as an integral over the Feynman
region of $\m$ (see (\ref{Gg})):
\begin{equation}
  \Gamma_4^{\rm {Feyn}}={2\over\pi}\int_{\F_{0,4}}\lf|\g(s,\,t,\,u)\rt|^2.
  \label{part-amp}
\end{equation}
Instead of integrating over the Feynman region we can integrate over three
unit disks $|\e|<1$, one for each channel. Consider for example the
contribution of the $u$ channel:
\begin{equation}
  \Gamma^{(u)}_4(s,\,t,\,u) = {2\over\pi}\int_{\F_u}|\g(s,\,t,\,u)|^2.
  \label{gammas}
\end{equation}
We can find $\g(s,\,t,\,u)$ from equation (\ref{kobastu}) which we rewrite
as
\begin{equation}
  \label{gstu}
  \g(s,\,t,\,u)={\l'(\e)\d\e\over
   \l(\e)^{{m^2\over2}-s}(1-\l(\e))^{{m^2\over2}-t}\chi^{{m^2\over2}+2} }.
\end{equation}
In terms of $\e$ the region of integration $\F_u$ is just the unit disk
$|\e|<1$.  Recall that (see eqn. (\ref{lamqa}))
\begin{equation}
\l(\e)= {3^3\over 2^4}\,\e^{-1} + {1\over 2} - {11\,\e\over 2^2\cdot3^3} -
   {1621\over 2^3\cdot 3^8}\,\e^3 - {413941\over 2\cdot3^{15}}\,\e^5 +
   \O(\e^7),
  \label{le}
\end{equation}
is of order of $\e^{-1}$ for small $\e$ and $\chi_u=\O(1)$.  Therefore we
can represent $\g(s,t,u)$ in the $u$ channel as
\begin{equation}
\g(s,t,u)= \e^{-2-{u\over2}}\sum_{n=0}^\infty c_n(s,t)\e^n\,\d\e.
  \label{gsexpand}
\end{equation}
We can now evaluate the integral (\ref{gammas}). If the coefficients $c_n$
vanish sufficiently fast for $n\to\infty$, this integral converges for
$\re\,u<-2$ and is given by
\begin{equation}
\Gamma^{(u)}_4(s,\,t,\,u) = \sum_{n=0}^\infty {4\,|c_n(s,t)|^2\over2n-2-u}.
  \label{Gsum}
\end{equation}
Note that $\pi$ from the prefactor in (\ref{gammas}) cancels with the area
of the unit disk.  Equation (\ref{Gsum}) shows that the amplitude has an
analytic continuation to the whole region $\re\,u>-2$, except for even
integer values of $u$, where it has first order poles. These poles
correspond to the spectrum of the closed string.

In order to find the constants $c_n(s, t)$ it is sufficient to find the
series expansion for $\gamma(s, t, 0)$. For the tachyonic potential we need
only $c_n(0,0)$, so let us restrict ourselves to this case,
\begin{equation}
\g^{(0)}=\g(0,\,0,\,0)=\chi_u^2\,\d\l.
  \label{gst0}
\end{equation}
First of all, recall that $\chi_u$ is an even function of $\e$ and that
$\l(-\e)=1-\l(\e)$. Indeed, when we make a twist by $\pi$ it is equivalent
to an exchange of the poles $z_2$ and $z_3$. This exchange does not affect
$\chi_u=\chi_{14}$ but changes $\l$ to $1-\l$. We can now conclude that
$\g^{(0)}$ is even with respect to $\e$ and $c_n(0,0)=0$ for odd $n$. In
particular, this means that massless states ($n=1$) are decoupled from the
tachyon. The sum over massive states which appears in eqn. (\ref{v4effd})
is given by
\begin{equation}
\sum_X\massive={3}\,\sum_{k=1}^\infty{2\,|c_{2k}(0,0)|^2\over 2\,k-1},
  \label{xsum}
\end{equation}
where we have introduced an extra factor of $3$ which comes from summation
over three channels.  Each term in the series corresponds to a particular
mass level and can be found by summing corresponding Feynman diagrams. For
example, on the lowest mass level there is only one state --- the tachyon,
and we therefore conclude that
\begin{equation}
{4\,|c_0|^2\over -2} = \tach = {\v_3^2\over -2}, \qquad\mbox{or}\qquad c_0
 = \half\v_3^2={3^9\over2^{12}}.
  \label{fromv3}
\end{equation}
One can similarly evaluate the Feynman diagrams for some other massive
levels and thus evaluate some more $c_n$. An alternative way to do this is
to use the series for $\chi_u$ and $\l$ from sect.~\ref{Infi} (see
eqn.(\ref{chiqa})) and evaluate $ \g^{(0)}$ directly as
\begin{equation}
 \g^{(0)}=\chi^2_u\d\l=\lf({3^{9}\over 2^{12}}\,\e^{-2} + {1377\over
   2^{10}} + {1399\over 2^{11}}\,\e^2 + {4504241\over 2^{9}\cdot3^9}\,\e^4
   + \O(\e^6) \rt)\d\e.
  \label{gappr}
\end{equation}

Although we can in principle find as many coefficients $c_n$ as we want, it
is very inefficient to evaluate $\v_4$ summing the series because it
converges very slowly. The reason for this poor convergence is that the
series for $\g^{(0)}$ diverges at $\e=1$. Indeed $\e=1$ corresponds to
$\b=0$ and we can use approximate formulae for $\l(\e)$ and $\chi_u(\e)$ in
the vicinity of this point to get:
\begin{equation}
 \g^{(0)} = \lf(8\,\ln\lf(8\pi\over1-\e\rt)+\O(1-\e)\rt)\d\e.
  \label{e1}
\end{equation}
Looking at the first term of this expansion we conclude that
\begin{equation}
  c_n\sim{1\over n}\qquad\mbox{for}\qquad n\to\infty
  \label{cninf}
\end{equation}
Therefore the series in (\ref{xsum}) converges as slowly as $\sum n^{-3}$.

Instead of summing the series we therefore decide to calculate the integral
itself. First of all, we have to regularize $\g^{(0)}$ by subtracting the
divergent term $(c_0/\e^{2})\d\e$. We can then evaluate convergent integral
numerically:
\begin{equation}
\sum_X\massive={6\over\pi}\int_{|\e|<1}
               \lf|\g^{(0)}-{c_0\over\e^{2}}\,\d\e\rt|^2 \approx6.011.
  \label{xsumn}
\end{equation}
\underline{Historical remarks.} Calculations of the Feynman region
contribution to the closed string amplitude are very similar to those in
the case of the open string. Indeed in the open string we have to consider
the same differential form $\g$ and integrate it along the real interval
$[-1, 1]$ in order to get a contribution from one channel. The results of
this section have been found in the realm of open string in the works of
Kosteleck\'y and Samuel \cite{ks2,ks1}.  Using different methods to those
applied here, they were able to find the quartic term in the effective
potential. The series expansion analogous to (\ref{gappr}) has been found
in Ref.~\cite{sloan} up to order $\e^2$ and it was verified that the
coefficients agree with what one gets from the Feynman diagrams with an
intermediate massive state.

\section{Bare four tachyon coupling constant and full effective
  potential}
\label{Bare}
\setcounter{equation}{0}

As we saw in the previous section the four punctured spheres which can be
obtained from Feynman diagrams do not cover the moduli space $\m$. The
contribution of the rest of $\m$ can be introduced in the string field
theory as elementary $4$-string coupling In this section we evaluate this
elementary coupling for the case of four tachyons.
\setlength{\unitlength}{1in}
\begin{figure}[t]
  \begin{center}
    \leavevmode
    \begin{picture}(5,4)
      \put(0,-0.2){\psfig{figure=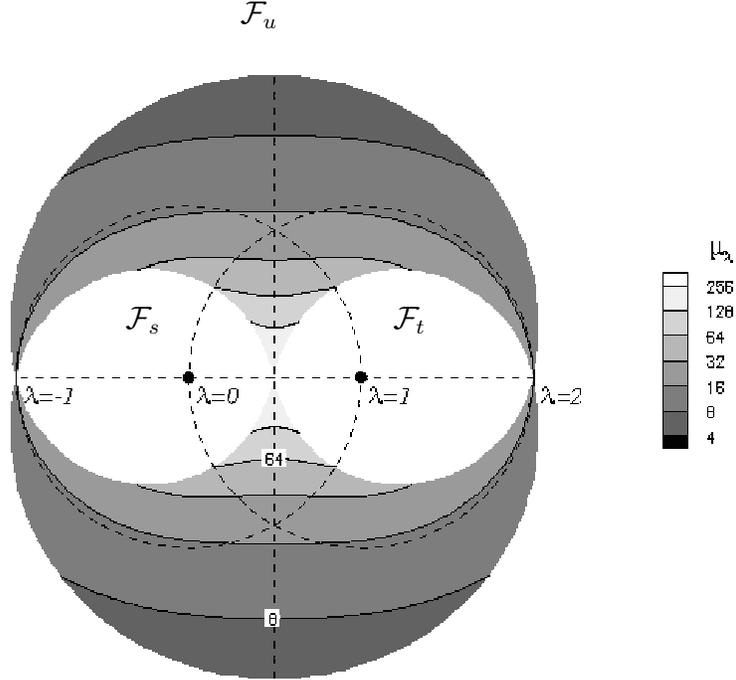,width=\the\textwidth}}
      \put(1.3,2.0){${\cal F}_s$} \put(2.7,2.0){${\cal F}_t$}
      \put(1.9,3.6){${\cal F}_u$}
    \end{picture}
  \end{center}
  \caption{The moduli space $\m$ and the measure of integration $\mu_\l$}
  \label{fig:lam}
\end{figure}

The four-string vertex $\V_{0,4}=\m\backslash\F_{0,4}$ can be easily
described in terms of the integral invariants $a$, $b$ and $c$ introduced
in sect.~\ref{Quad}. The whole moduli space can be parameterized by real
values of these invariants varying from $0$ to $2\pi$ restricted by the
condition $a+b+c=2\pi$.  In fact, each triple defines two points $\l$ and
$\bar\l$ in $\m$, so we need two copies of the $abc$ triangle to cover
$\m$. The four-string vertex can now be described as a region in the $abc$
triangle defined by
\begin{equation}
  a>\pi,\quad b>\pi, \quad\mbox{and}\quad c>\pi.
  \label{v4abc}
\end{equation}
The four tachyon coupling is given by the same integral as the amplitude,
but taken not over the whole moduli space, rather restricted to only
$\V_{0,4}$.
\begin{equation}
\v_4 ={2\over\pi}\int_{\V_{0,4}}|\chi_u|^4\d^2\l,
  \label{v4int}
\end{equation}
where $\d^2\l =\d\im\,\l\,\d\re\,\l$.

\noindent \underline{Numerical results.}
\begin{figure}[ht]
  \begin{center}
    \leavevmode \psfig{figure=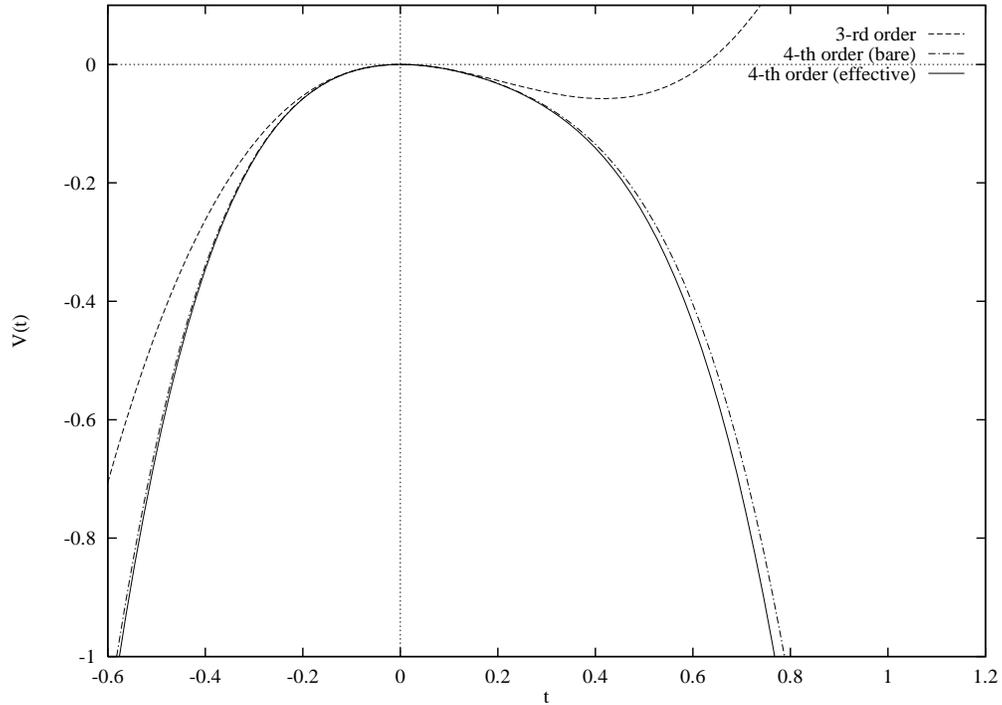,width=\textwidth}
  \end{center}
  \caption{Tachyonic potential.}
  \label{fig:pot}
\end{figure}
For the numerical calculations we use the complex secant method with the
starting point given by (\ref{qt3}) in order to solve the main
equation. Then we calculate $\chi_u$ and $\lp$ using the first few terms of
the infinite products (\ref{chipro}) and (\ref{lampro}).  Results are
presented in Fig.~\ref{fig:lam} which shows the region of integration
$\V_{0,4}$ and the contour plot of the measure $\mu_\l=|\chi_u|^4$. We
perform calculations only for $\dl\geq0$, $\b\leq1/3-(2/3\pi)\,\dl$ which
is $1/6$ of the whole $abc$ triangle (see Fig.~\ref{fig:abc}).  The values
of $\lp$ and $\mu_\l$ in the rest of the triangle are found from
symmetry. As we can see $\mu_\l$ has its maximum value of $2^8=256$ at
$\l=1/2$ and drops exponentially as we go away from this point.  Note, that
the value of the measure $\mu_\l$ at the point where the unit circle
intersects the boundary of $\V_{0,4}$ is equal to $64$ exactly (at least up
to machine precision $10^{-10}$).  We could not find any explanation to
this fact.

We have performed numerical integration triangulating $1/12$-th of
$\V_{0,4}$ correspondent to $\dl\geq0$ and $\b\leq1/3-(2/3\pi)\,\dl$. Here
we present the result of the calculation which involved about $500,000$
triangles.
\begin{equation}
  \fbox{$\ds\v_4={2\over\pi}\int_{\V_{0,4}}|\chi_u|^4\d^2\l \approx 72.39$}
    % 113.71 * 2 / pi
  \label{v4-value}
\end{equation}
Combining (\ref{v4-value}) and (\ref{xsumn}) we can finally write the
tachyonic potential up to the fourth order:
\begin{equation}
  V^{\rm eff}(t) = -t^2 + 1.60181\,t^3-3.267\,t^4+\O(t^5).
  \label{v4num}
\end{equation}
We present the plot of the effective tachyonic potential computed up to the
fourth term in Fig.~\ref{fig:pot}. One can see that the fourth order term
is big enough to destroy the local minimum suggested by the third order
approximation (dashed line in the plot). The plot also shows the bare
tachyonic potential computed up to the fourth order. One can see that the
effective four-tachyon interaction gives only a small correction.

\section*{Acknowledgments}
I would like to thank B.~Zwiebach for encouraging me to solve this problem
and for numerous helpful discussions. I also thank J.~Keiper from WRI for
providing me with a {\it Mathematica}$^{\rm (R)}$ package for calculating
the elliptic functions.

\end{document}